# Ranking Large Temporal Data

Jeffrey Jestes, Jeff M. Phillips, Feifei Li, Mingwang Tang
*School of Computing, University of Utah, Salt Lake City, USA*
{jestes, jeffp, lifeifei, tang}@cs.utah.edu

Ranking temporal data has not been studied until recently, even though ranking is an important operator (being promoted as a first-class citizen) in database systems. However, only the *instant* top-$k$ queries on temporal data were studied in, where objects with the $k$ highest scores at a query time instance $t$ are to be retrieved. The instant top-$k$ definition clearly comes with limitations (sensitive to outliers, difficult to choose a meaningful query time $t$). A more flexible and general ranking operation is to rank objects based on the aggregation of their scores in a query interval, which we dub the *aggregate* top-$k$ query on temporal data. For example, return the top-10 weather stations having the highest average temperature from 10/01/2010 to 10/07/2010; find the top-20 stocks having the largest total transaction volumes from 02/05/2011 to 02/07/2011. This work presents a comprehensive study to this problem by designing both exact and approximate methods (with approximation quality guarantees). We also provide theoretical analysis on the construction cost, the index size, the update and the query costs of each approach. Extensive experiments on large real datasets clearly demonstrate the efficiency, the effectiveness, and the scalability of our methods compared to the baseline methods.

## 1. INTRODUCTION

Temporal data has important applications in numerous domains, such as in the financial market, in scientific applications, and in the biomedical field. Despite the extensive literature on storing, processing, and querying temporal data, and the importance of ranking (which is considered as a first-class citizen in database systems [9]), ranking temporal data has not been studied until recently [15]. However, only the *instant* top-$k$ queries on temporal data were studied in [15], where objects with the $k$ highest scores at a query time instance $t$ are to be retrieved; it was denoted as the top-$k(t)$ query in [15]. The instant top-$k$ definition clearly comes with obvious limitations (sensitivity to outliers, difficulty in choosing a meaningful single query time $t$). A much more flexible and general ranking operation is to rank temporal objects based on the aggregation of their scores in a query interval, which we dub the *aggregate* top-$k$ query on temporal data, or top-$k(t_1, t_2, \sigma)$ for an interval $[t_1, t_2]$ and an aggregation function $\sigma$. For example, return the top-10 weather stations having the highest average temperature from 10/01/2010 to 10/07/2010; find the top-20 stocks having the largest total transaction volumes from 02/05/2011 to 02/07/2011.

Clearly, the instant top-$k$ query is a special case of the aggregate top-$k$ query (when $t_1 = t_2$). The work in [15] shows that even the instant top-$k$ query is hard!

**Problem formulation.** In temporal data, each object has at least one score attribute $A$ whose value changes over time, e.g., the temperature readings in a sensor database. An example of real temperature data from the MesoWest project appears in Figure 1. In general, we can represent the score attribute $A$ of an object as an arbitrary function $f : \mathbb{R} \rightarrow \mathbb{R}$ (time to score), but for arbitrary temporal data, $f$ could be expensive to describe and process. In practice, applications often approximate $f$ using a piecewise linear function $g$ [6, 1, 12, 11]. The problem of approximating an arbitrary function $f$ by a piecewise linear function $g$ has been extensively studied (see [12, 16, 6, 1] and references therein). Key observations are: 1) more segments in $g$ lead to better approximation quality, but also are more expensive to represent; 2) adaptive methods, by allocating more segments to regions of high volatility and less to smoother regions, are better than non-adaptive methods with a fixed segmentation interval.

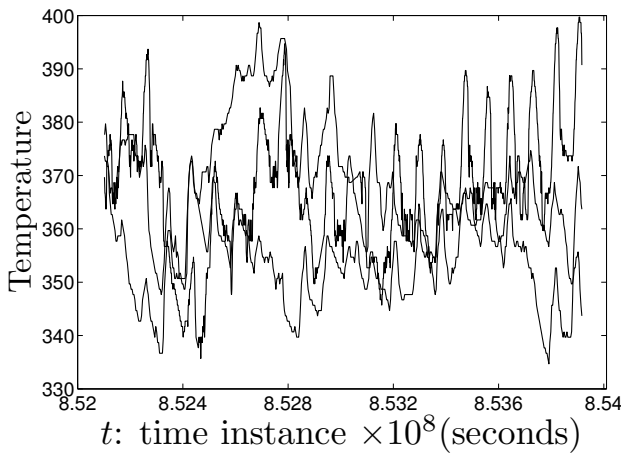

**Figure 1: MesoWest data.**

In this paper, for the ease of discussion and illustration, we focus on temporal data represented by piecewise linear functions. Nevertheless, our results can be extended to other representations of time series data, as we will discuss in Section 4. Note that a lot of work in processing temporal data also assumes the use of piecewise linear functions as the main representation of the temporal data [6, 1, 12, 11, 14], including the prior work on the instant top-$k$ queries in temporal data [15]. That said, how to approximate $f$ with $g$ is beyond the scope of this paper, and we assume that the data has already been converted to a piecewise linear representation by *any* segmentation method. In particular, we require *neither* them having the same number of segments *nor* them having the aligned starting/ending time instances for segments from different functions. Thus it is possible that the data is collected from a variety of sources after each applying different preprocessing modules.

That said, formally, there are $m$ objects in a temporal database; the $i$th object $o_i$ is represented by a piecewise linear function $g_i$ with $n_i$ number of (linear line) segments. There are a total of $N = \sum_{i=1}^{m} n_i$ segments from all objects. The temporal range of any object is in $[0, T]$. An aggregate top-$k$ query is denoted as top-$k(t_1, t_2, \sigma)$ for some aggregation function $\sigma$, which is to retrieve the $k$ objects with the $k$ highest aggregate scores in the range $[t_1, t_2]$, denoted as an ordered set $\mathcal{A}(k, t_1, t_2)$ (or simply $\mathcal{A}$ when the context is clear). The aggregate score of $o_i$ in $[t_1, t_2]$ is defined as $\sigma(g_i(t_1, t_2))$, or simply $\sigma_i(t_1, t_2)$, where $g_i(t_1, t_2)$ denotes the set of all possible values of function $g_i$ evaluated at every time instance in $[t_1, t_2]$ (clearly an infinite set for continuous time domain). For example, when $\sigma = \text{sum}$, the aggregate score for $o_i$ in $[t_1, t_2]$ is $\int_{t_1}^{t_2} g_i(t)dt$. An example of a sum top-2 query is shown in Figure 2, and its answer is $\{o_3, o_1\}$.

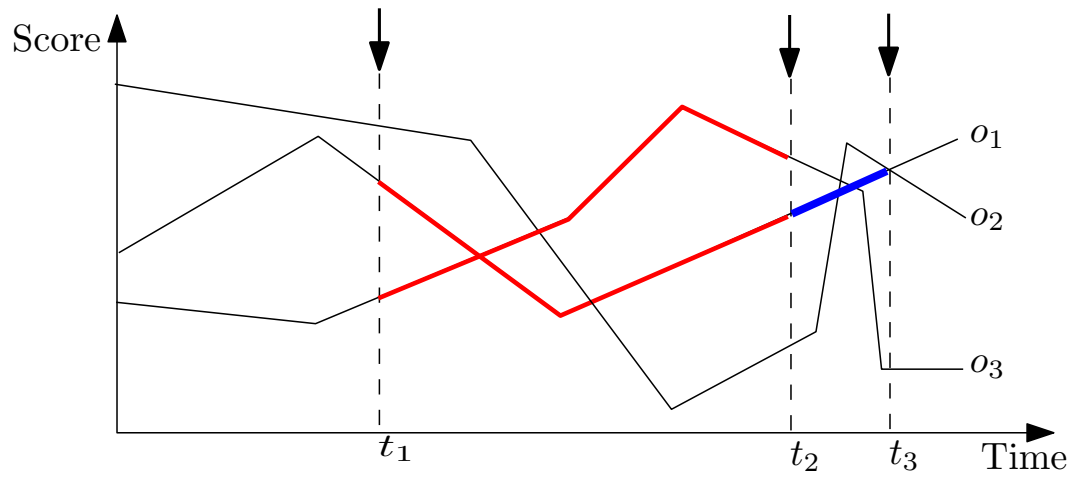

**Figure 2: A top-$2(t_1, t_2, \text{sum})$ query example.**

For ease of illustration, we assume non-negative scores by default. This restriction is removed in Section 4. We also assume a max possible value $k_{\max}$ for $k$.

**Our contributions.** A straightforward observation is that a solution to the instant top-$k$ query cannot be directly applied to solve the aggregate top-$k$ query since: 1) the temporal dimension can be continuous; and 2) an object might not be in the top-$k$ set for any top-$k(t)$ query for $t \in [t_1, t_2]$, but still belong to $\mathcal{A}(k, t_1, t_2)$ (for example, $\mathcal{A}(1, t_2, t_3)$ in Figure 2 is $\{o_1\}$, even though $o_1$ is never a top-$1(t)$ object for any $t \in [t_2, t_3]$). Hence, the trivial solution (denoted as EXACT1) is for each query to compute $\sigma_i(t_1, t_2)$ of every object and insert them into a priority queue of size $k$, which takes $O(m(N + \log k))$ time per query and is clearly not scalable for large datasets (although our implementation slightly improves this query time as described in Section 2). Our goal is then to design IO and computation efficient algorithms which can outperform the trivial solution and work well regardless if data fits in main memory or not. A design principle we have followed is to leverage on existing indexing structures whenever possible (so these algorithms can be easily adopted in practice). Our work focuses specifically on $\sigma = \text{sum}$, and we make the following contributions:

- We design a novel exact method in Section 2, based on using a single interval tree (EXACT3).
- We present two approximate methods (and several variants) in Section 3. Each offers an approximation $\widetilde{\sigma}_i(t_1, t_2)$ on the aggregate score $\sigma_i(t_1, t_2)$ for objects in any query interval. We say $\tilde{X}$ is an $(\varepsilon, \alpha)$-approximation of $X$ if $X/\alpha - \varepsilon M \leq \tilde{X} \leq X + \varepsilon M$ for user-defined parameters $\alpha \geq 1$, $\varepsilon > 0$ and where $M = \sum_{i=1}^{m} \sigma_i(0, T)$. Now, for $i \in [1, m]$, $[t_1, t_2] \subseteq [0, T]$, the APPX1 method guarantees that $\widetilde{\sigma}_i(t_1, t_2)$ is an $(\varepsilon, 1)$-approximation of $\sigma_i(t_1, t_2)$, and the APPX2 method guarantees $\widetilde{\sigma}_i(t_1, t_2)$ is an $(\varepsilon, 2\log(1/\varepsilon))$-approximation of $\sigma_i(t_1, t_2)$. We show an $(\varepsilon, \alpha)$-approximation on $\sigma_i(t_1, t_2)$ implies an approximation $\widetilde{\mathcal{A}}(k, t_1, t_2)$ of $\mathcal{A}(k, t_1, t_2)$ such that the aggregate score of the $j$th ranked ($1 \leq j \leq k$) object in $\widetilde{\mathcal{A}}(k, t_1, t_2)$ is always an $(\varepsilon, \alpha)$-approximation of the aggregate score of the $j$th ranked object in $\mathcal{A}(k, t_1, t_2)$.
- We extend our results to general functions $f$ for temporal data, other possible aggregates, negative scores, and deal with updates in Section 4.
- We show extensive experiments on massive real data sets in Section 5. The results clearly demonstrate the efficiency, effectiveness and scalability of our methods compared to the baseline. Our approximate methods are especially appealing when approximation is admissible, given their better query costs than exact methods and high quality approximations.

| Symbol | Description |
|---|---|
| $\mathcal{A}(k, t_1, t_2)$ | ordered top-$k$ objects for top-$k(t_1, t_2, \sigma)$. |
| $\widetilde{\mathcal{A}}(k, t_1, t_2)$ | an approximation of $\mathcal{A}(k, t_1, t_2)$. |
| $\mathcal{A}(j), \widetilde{\mathcal{A}}(j)$ | the $j$th ranked object in $\mathcal{A}$ or $\widetilde{\mathcal{A}}$. |
| $B$ | block size. |
| $\mathcal{B}$ | set of breakpoints ($\mathcal{B}_1$ and $\mathcal{B}_2$ are special cases). |
| $\mathcal{B}(t)$ | smallest breakpoint in $\mathcal{B}$ larger than $t$. |
| $g_i$ | the piecewise linear function of $o_i$. |
| $g_{i,j}$ | the $j$th line segment in $g_i$, $j \in [1, n_i]$. |
| $g_i(t_1, t_2)$ | the set of all possible values of $g_i$ in $[t_1, t_2]$. |
| $k_{\max}$ | the maximum $k$ value for user queries. |
| $\ell(t)$ | the value of a line segment $\ell$ at time instance $t$. |
| $m$ | total number of objects. |
| $M$ | $M = \sum_{i=1}^{m} \sigma_i(0, T)$. |
| $n_i$ | number of line segments in $g_i$. |
| $n, n_{\text{avg}}$ | $\max\{n_1, n_2, \ldots, n_m\}$, $\text{avg}\{n_1, n_2, \ldots, n_m\}$ |
| $N$ | number of line segments of all objects. |
| $o_i$ | the $i$th object in the database. |
| $q_i$ | number of segments in $g_i$ overlapping $[t_1, t_2]$. |
| $r$ | number of breakpoints in $\mathcal{B}$, bounded $O(1/\varepsilon)$. |
| $(t_{i,j}, v_{i,j})$ | $j$th end-point of segments in $g_i$, $j \in [0, n_i]$. |
| $\sigma_i(t_1, t_2)$ | aggregate score of $o_i$ in an interval $[t_1, t_2]$. |
| $\widetilde{\sigma}_i(t_1, t_2)$ | an approximation of $\sigma_i(t_1, t_2)$. |
| $[0, T]$ | the temporal domain of all objects. |

**Table 1: Frequently used notations.**

We survey the related work in Section 6, and conclude in Section 7. Table 1 summarizes our notations. Figure 3 summarizes the upper bounds on the preprocessing cost, the index size, the query cost, the update cost, and the approximation guarantee of all methods.

## 2. EXACT METHODS

As explained in Section 1, a trivial exact solution EXACT1 is to find the aggregate score of each object in the query interval and insert them into a priority queue of size $k$. We can improve this approach by indexing line segments from all objects with a B+-tree, where the key for a data entry $e$ is the value of the time-instance for the left-end point of a line segment $\ell$, and the value of $e$ is just $\ell$. Given a query interval $[t_1, t_2]$, this B+-tree allows us to find all line segments that contains $t_1$ in $O(\log_B N)$ IOs. A sequential scan (till $t_2$) then can retrieve all line segments whose temporal dimensions overlap with $[t_1, t_2]$ (either fully or partially). In this process, we simply maintain $m$ running sums, one per object in the database. Suppose the $i$th running sum of object $o_i$ is $s_i$ and it is initialized with the value 0. Given a line segment $\ell$ defined by $(t_{i,j}, v_{i,j})$ and $(t_{i,j+1}, v_{i,j+1})$ from $o_i$ (see an example in Figure 4), we define an interval $I = [t_1, t_2] \cap [t_{i,j}, t_{i,j+1}]$, let $t_L = \max\{t_1, t_{i,j}\}$ and $t_R = \min\{t_2, t_{i,j+1}\}$, and update $s_i = s_i + \sigma_i(I)$, where

$$\sigma_i(I) = \begin{cases} 0, \text{ if } t_2 < t_L \text{ or } t_1 > t_R; \\ \frac{1}{2}(t_R - t_L)(\ell(t_R) + \ell(t_L)), \quad \text{else.} \end{cases} \tag{1}$$

Note that $\ell(t)$ is the value of the line segment $\ell$ at time $t$. Note that if we follow the sequential scan process described above, we will only deal with line segments that do overlap with the temporal range $[t_1, t_2]$, in which the increment to $s_i$ corresponds to the second case in (1). It is essentially an integral from $t_L = \max\{t_1, t_{i,j}\}$ to $t_R = \min\{t_2, t_{i,j+1}\}$ w.r.t. $\ell$, i.e., $\int_{t_L}^{t_R} \ell(t)dt$. This range $[t_L, t_R]$ of $\ell$ also defines a trapezoid, hence, it is equal to the area of this trapezoid, which yields the formula in (1).

When we have scanned all line segments up to $t_2$ from the B+-tree, we stop and assign $\sigma_i(t_1, t_2) = s_i$ for $i = 1$ to $m$. Finally, we insert $(i, \sigma_i(t_1, t_2))$, for $i = 1$ to $m$, into a priority queue of size $k$ sorted in the descending order of $\sigma_i(t_1, t_2)$. The answer $\mathcal{A}(k, t_1, t_2)$ is the (ordered) object ids in this queue when the last pair $(m, \sigma_m(t_1, t_2))$ has been processed.

| | index size | construction cost | query cost | update cost | approximation |
|---|---|---|---|---|---|
| EXACT1 | $O(\frac{N}{B})$ | $O(\frac{N}{B}\log_B N)$ | $O(\log_B N + \frac{\sum_{i=1}^m q_i}{B})$ | $O(\log_B N)$ | $(0, 1)$ |
| EXACT2 | $O(\frac{N}{B})$ | $O(\sum_{i=1}^m \frac{n_i}{B}\log_B n_i)$ | $O(\sum_{i=1}^m \log_B n_i)$ | $O(\log_B n)$ | $(0, 1)$ |
| EXACT3 | $O(\frac{N}{B})$ | $O(\frac{N}{B}\log_B N)$ | $O(\log_B N + \frac{m}{B})$ | $O(\log_B N)$ | $(0, 1)$ |
| APPX1 | $O(\frac{r^2}{B}k_{\max})$ | $O(\frac{N}{B}(\log_B N + r))$ | $O(\frac{k}{B} + \log_B r)$ | $O(\frac{1}{B}(\log_B N + r))$ | $(\varepsilon, 1)$ |
| APPX2 | $O(\frac{r}{B}k_{\max})$ | $O(\frac{N}{B}(\log_B N + \log r))$ | $O(k\log r)$ | $O(\frac{1}{B}(\log_B N + \log r))$ | $(\varepsilon, 2\log r)$ |

**Figure 3: IO costs, with block size $B$; for simplicity, $\log_B k_{\max}$ terms are absorbed in $O(\cdot)$ notation.**

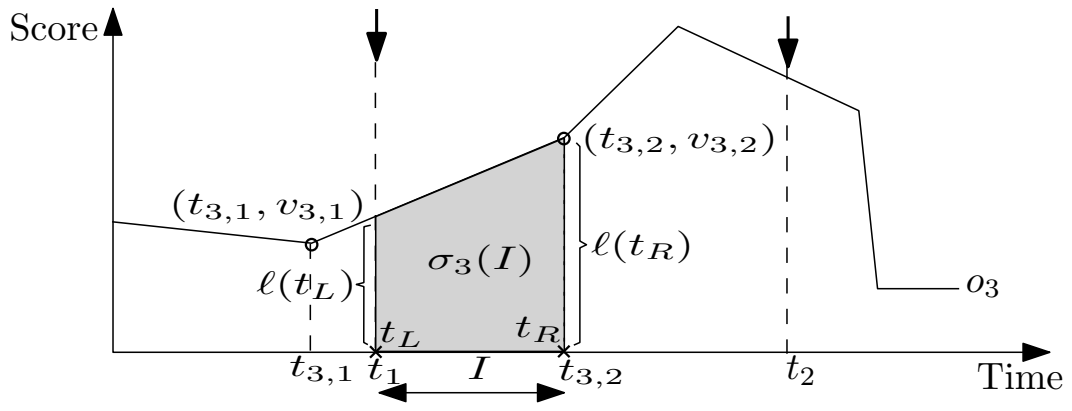

**Figure 4: Compute $\sigma_i([t_1, t_2] \cap [t_{i,j}, t_{i,j+1}])$.**

This method EXACT1 has a cost of $O((N/B)\log_B N)$ IOs for building the B+-tree, an index size of $O(N/B)$ blocks, and a query cost of $O(\log_B N + \sum_{i=1}^m q_i/B + (m/B)\log_B k)$ IOs where $q_i$ is the number of line segments from $o_i$ overlapping with the temporal range $[t_1, t_2]$ of a query $q$=top-$k(t_1, t_2, \text{sum})$. In the worst case, $q_i = n_i$ for each $i$, then the query cost becomes $O(N/B)$!

**A forest of B+-trees.** EXACT1 becomes quite expensive when there are a lot of line segments in $[t_1, t_2]$, and its asymptotic query cost is actually $O(N/B)$ IOs, which is clearly non-scalable. The bottleneck of EXACT1 is the computation of the aggregate score of each object. One straight forward idea to improve the aggregate score computation is to leverage on precomputed prefix-sums [7]. We apply the notion of prefix-sums to continuous temporal data by precomputing the aggregate scores of some selected intervals in each object; this preprocessing helps reduce the cost of computing the aggregate score for an arbitrary interval in an object. Let $(t_{i,j}, v_{i,j})$ be the $j$th end-point of segments in $g_i$, where $j \in \{0, \dots, n_i\}$; clearly, the $j$th segment in $g_i$ is then defined by $((t_{i,j-1}, v_{i,j-1}), (t_{i,j}, v_{i,j}))$ for $j \in \{1, \dots, n_i\}$, which we denote as $g_{i,j}$. Then define intervals $I_{i,\ell} = [t_{i,0}, t_{i,\ell}]$ for $\ell = 1, \dots, n_i$, and compute the aggregate score $\sigma_i(I_{i,\ell})$ for each.

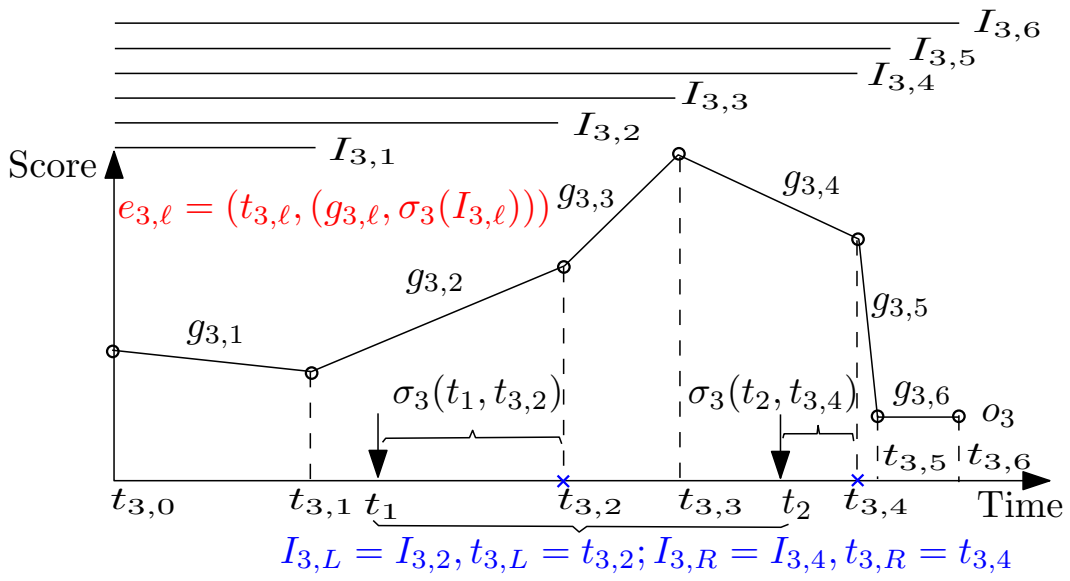

**Figure 5: The method EXACT2.**

Once we have $(I_{i,\ell}, \sigma_i(I_{i,\ell}))$s, we build a B+-tree to index them. Specifically, we make a leaf-level data entry $e_{i,\ell}$ for $(I_{i,\ell}, \sigma_i(I_{i,\ell}))$, where the key in $e_{i,\ell}$ is $t_{i,\ell}$ (the right end-point of $I_{i,\ell}$), and the value of $e_{i,\ell}$ includes both $g_{i,\ell}$ and $\sigma_i(I_{i,\ell})$. Given $\{e_{i,1}, \dots, e_{i,n_i}\}$ for $o_i$, we bulk-load a B+-tree $T_i$ using them as the leaf-level data entries (see Figure 5 for an example).

We do this for each object, resulting in $m$ B+-trees. Given $T_i$, we can compute $g_i(t_1, t_2)$ for any interval $[t_1, t_2]$ efficiently. We first find the data entry $e_{i,L}$ such that its key value $t_{i,L}$ is the first succeeding key value of $t_1$; we then find the data entry $e_{i,R}$ such that its key value $t_{i,R}$ is the first succeeding key value of $t_2$. Next, we can calculate $\sigma_i(t_1, t_{i,L})$ using $g_{i,L}$ (stored in $e_{i,L}$), and $\sigma_i(t_2, t_{i,R})$ using $g_{i,R}$ (stored in $e_{i,R}$), simply based on (1). Finally,

$$\sigma_i(t_1, t_2) = \sigma_i(I_{i,R}) - \sigma_i(I_{i,L}) + \sigma_i(t_1, t_{i,L}) - \sigma_i(t_2, t_{i,R}), \quad (2)$$

where $\sigma_i(I_{i,R})$, $\sigma_i(I_{i,L})$ are available in $e_{i,R}$, $e_{i,L}$ respectively. Figure 5 also gives a query example using $o_3$.

Once all $\sigma_i(t_1, t_2)$'s are computed for $i = 1, \dots, m$, the last step is the same as that in EXACT1.

We denote this method as EXACT2. Finding $e_{i,L}$ and $e_{i,R}$ from $T_i$ takes only $\log_B n_i$ cost, and calculating (2) takes $O(1)$ time. Hence, its query cost is $O(\sum_{i=1}^m \log_B n_i + m/B\log_B k)$ IOs. The index size of this method is the size of all B+-trees, where $T_i$'s size is linear to $n_i$; so the total size is $O(N/B)$ blocks. Note that computing $\{\sigma_i(I_{i,1}), \dots, \sigma_i(I_{i,n_i})\}$ can be easily done in $O(n_i/B)$ IOs, by sweeping through the line segments in $g_i$ sequentially from left to right, and using (1) incrementally (i.e., computing $\sigma_i(I_{i,\ell+1})$ by initializing its value to $\sigma_i(I_{i,\ell})$). Hence, the construction cost is dominated by building each tree $T_i$ with cost $O((n_i/B)\log_B n_i)$. The total construction cost is $O(\sum_{i=1}^m (n_i/B)\log_B n_i)$.

**Using one interval tree.** When $m$ is large (as is the case for the real data sets we explore in Section 5), querying $m$ B+-trees becomes very expensive, partly due to the overhead of opening and closing $m$ disk files storing these B+-trees. Hence, an important improvement is to somehow index the data entries from all $m$ B+-trees in a single disk-based data structure.

Consider any object $o_i$, let intervals $I_{i,1}, \dots, I_{i,n_i}$ be the same as that in EXACT2, where $I_{i,\ell} = [t_{i,0}, t_{i,\ell}]$. Furthermore, we define intervals $I^-_{i,1}, \dots, I^-_{i,n_i}$, such that $I^-_{i,\ell} = I_{i,\ell} - I_{i,\ell-1}$ (let $I_{i,0} = [t_{i,0}, t_{i,0}]$), i.e., $I^-_{i,\ell} = [t_{i,\ell-1}, t_{i,\ell}]$.

Next, we define a data entry $e_{i,\ell}$ such that its key is $I^-_{i,\ell}$, and its value is $(g_{i,\ell}, \sigma_i(I_{i,\ell}))$, for $\ell = 1, \dots, n_i$. Clearly, an object $o_i$ yields $n_i$ such data entries. Figure 6 illustrates an example using the same setup in Figure 5. When we collect all such entries from all objects, we end up with $N$ data entries in total. We denote these data entries as a set $I^-$; and it is interesting to note that the key value of each data entry in $I^-$ is an interval. Hence, we can index $I^-$ using a disk-based interval tree $S$ [13,4,3].

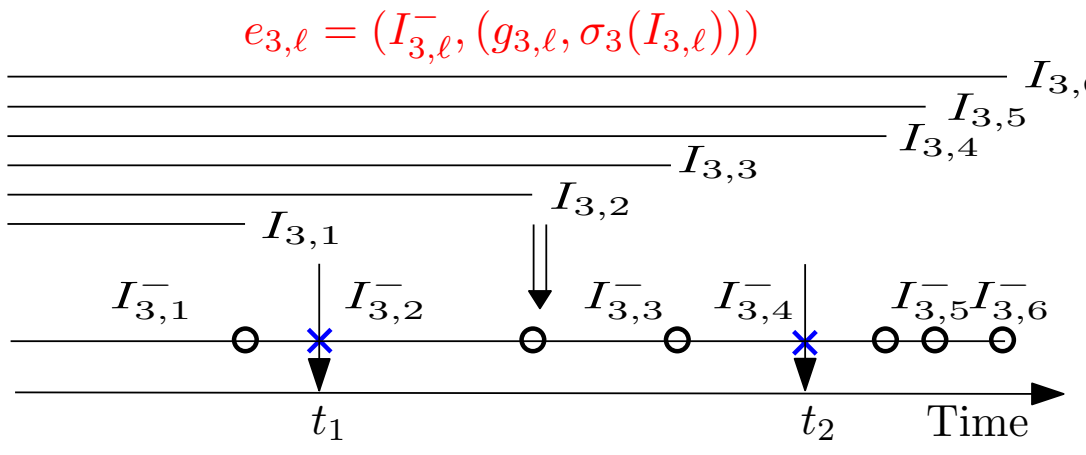

**Figure 6: The method EXACT3.**

Given this interval tree $S$, computing $\sigma_i(t_1, t_2)$ can now be reduced to two stabbing queries, using $t_1$ and $t_2$ respectively, which return the entries in $S$ whose key values (intervals in $I^-$) contain $t_1$ or $t_2$ respectively. Note that each such stabbing query returns exactly $m$ entries, one from each object $o_i$. This is because that: 1) any two intervals $I^-_{i,x}$, $I^-_{i,y}$ for $x \neq y$ from $o_i$ satisfies $I^-_{i,x} \cap I^-_{i,y} = \emptyset$; 2) and $I^-_{i,1} \cup I^-_{i,2} \cup \cdots \cup I^-_{i,n_i} = [0, T]$.

Now, suppose the stabbing query of $t_1$ returns an entry $e_{i,L}$ from $o_i$ in $S$, and the stabbing query of $t_2$ returns an entry $e_{i,R}$ from $o_i$ in $S$. It is easy to see that we can calculate $\sigma_i(t_1, t_2)$ just as (2) does in EXACT2 (see Figure 6). Note that using only these two stabbing queries are sufficient to compute all $\sigma_i(t_1, t_2)$'s for $i = 1, \ldots, m$.

Given $N$ data entries, the external interval tree has a linear size $O(N/B)$ blocks and takes $O((N/B)\log_B N)$ IOs to build [4] (building entries $\{e_{i,1}, \ldots, e_{i,n_i}\}$ for $o_i$ takes only $O(n_i/B)$ cost). The two stabbing queries take $O(\log_B N + m/B)$ IOs [4]; hence, the total query cost, by adding the cost of inserting $\sigma_i(t_1, t_2)$'s into a priority queue of size $k$, is $O(\log_B N + (m/B)\log_B k)$.

**Remarks.** One technique we do not consider is indexing temporal data with R-trees to solve aggregate top-$k$ queries. R-trees constructed over temporal data have been shown to perform orders of magnitude worse than other indexing techniques for answering instant top-$k$ queries, even when branch-and-bound methods are used [15]. Given this fact, we do not attempt to extend the use of R-trees to solve the harder aggregate top-$k$ query.

Temporal aggregation with range predicates has been studied in the classic work [22, 21], however, with completely different objectives. Firstly, they dealt with multi-versioned keys instead of time-series data, i.e., each key is alive with a constant value during a time period before it gets deleted. One can certainly model these keys as temporal objects with constant functions following our model (or even piecewise constant functions to model also updates to keys, instead of only insertions and deletions of keys). But more importantly, their definitions of the aggregation [22, 21] are fundamentally different from ours. The goal in [21] is to compute the sum of key values alive at a time instance, or alive at a time interval intersecting a query interval. The work in [22] extends [21] by allowing a range predicate on the key dimension as well, i.e., its goal is to compute the sum of key values that 1) are alive at a time instance, or alive at a time interval intersecting a query interval; 2) and are within a specified query range in the key dimension.

Clearly, these aggregations [22,21] are different from ours. They want to compute *a single aggregation* of all keys that "fall within" (are alive in) a two-dimensional query rectangle; while our goal is to compute the aggregate score values of many individual objects over a time interval (then rank objects based on these aggregations).

Zhang et al. [22] also extended their investigation to compute the sum of weighted key values, where each key value (that is alive in a two-dimensional query rectangle) is multiplied by a weight proportional to how long it is alive on the time dimension within the query interval. This weighted key value definition will be the same as our aggregation definition if an object's score is a constant in the query interval. They also claimed that their solutions can still work when the key value is not a constant, but a function with certain types of constraints. Nevertheless, even in these cases, their goal is to compute *a single sum over all weighted key values* for an arbitrary two-dimensional query rectangle, rather than each individual weighted key value over a time interval. Constructing $m$ such structures, a separate one for *each* of the $m$ objects in our problem, and only allowing an unbounded key domain can be seen as similar to our EXACT2 method, which on large data corpuses is the least efficient technique we consider. These fundamental differences make these works almost irrelevant in providing helpful insights for solving our temporal aggregation problems.

# 3. APPROXIMATE METHODS

The exact approaches require explicit computation of $\sigma_i(t_1, t_2)$ for each of $m$ objects, and we manage to reduce the IO cost of this from roughly $N/B$ to $m$ to $m/B$. Yet, on real data sets when $m$ is quite large, this can still be infeasible for fast queries. Hence we now study approximate methods that allow us to remove this requirement of computing all $m$ aggregates, while still allowing *any* query $[t_1, t_2]$ over the continuous time domain.

Our approximate methods focus on constructing a set of breakpoints $\mathcal{B} = \{b_1, b_2, \ldots, b_r\}, b_i \in [0, T]$ in the time domain, and snapping queries to align with these breakpoints. We prove the returned value $\widetilde{\sigma}_i(t_1, t_2)$ for any curve $(\varepsilon, 1)$-approximates $\sigma_i(t_1, t_2)$. The size of the breakpoints and time for queries will be independent of the total number of segments $N$ or objects $m$.

In this section we devise two methods for constructing $r$ breakpoints BREAKPOINTS1 and BREAKPOINTS2. The first method BREAKPOINTS1 guarantees $r = \Theta(1/\varepsilon)$ and is fairly straightforward to construct. The second method requires more advanced techniques to construct efficiently and guarantees $r = O(1/\varepsilon)$, but can be much smaller in practice.

Then given a set of breakpoints, we present two ways to answer approximate queries on them: QUERY1 and QUERY2. The first approach QUERY1 constructs $O(r^2)$ intervals, and uses a two-level B+-tree to retrieve the associated top $k$ objects list from the one interval snapped to by the query. The second approach QUERY2 only builds $O(r)$ intervals and their associated $k_{\max}$ top objects, and on a query narrows the list of possible top $k$-objects to a reduced set of $O(k \log r)$ objects. Figure 7 shows an outline of these methods.

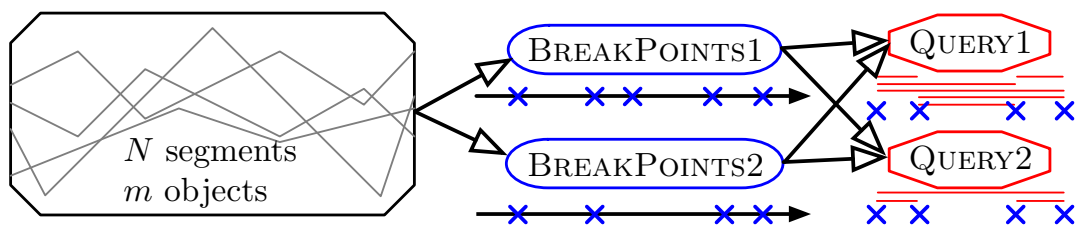

**Figure 7: Outline of approximate methods.**

We define the following approximation metrics.

**Definition 1** $G$ is an $(\varepsilon, \alpha)$-approximation algorithm of the aggregate scores if for any $i \in [1, m], [t_1, t_2] \subseteq [0, T]$, $G$ returns $\widetilde{\sigma}_i(t_1, t_2)$ such that $\sigma_i(t_1, t_2)/\alpha - \varepsilon M \leq \widetilde{\sigma}_i(t_1, t_2) \leq \sigma_i(t_1, t_2) + \varepsilon M$, for user-defined parameters $\alpha \geq 1, \varepsilon > 0$.

**Definition 2** For $\mathcal{A}(k, t_1, t_2)$ (or $\widetilde{\mathcal{A}}(k, t_1, t_2)$), let $\mathcal{A}(j)$ (or $\widetilde{\mathcal{A}}(j)$) be the $j$th ranked object in $\mathcal{A}$ (or $\widetilde{\mathcal{A}}$). $R$ is an $(\varepsilon, \alpha)$-approximation algorithm of top-$k(t_1, t_2, \sigma)$ queries if for any $k \in [1, k_{\max}], [t_1, t_2] \subseteq [0, T]$, $R$ returns $\widetilde{\mathcal{A}}(k, t_1, t_2)$ and $\widetilde{\sigma}_{\widetilde{\mathcal{A}}(j)}(t_1, t_2)$ for $j \in [1, k]$, s.t. $\widetilde{\sigma}_{\widetilde{\mathcal{A}}(j)}(t_1, t_2)$ is an $(\varepsilon, \alpha)$-approximation of $\sigma_{\widetilde{\mathcal{A}}(j)}(t_1, t_2)$ and $\sigma_{\mathcal{A}(j)}(t_1, t_2)$.

Definition 2 states that $\widetilde{\mathcal{A}}$ will be a good approximation of $\mathcal{A}$ if $(\varepsilon, \alpha)$ are small, since at each rank the two objects from $\widetilde{\mathcal{A}}$ and $\mathcal{A}$ respectively will have really close aggregate scores. This implies that the exact ranking order in $\mathcal{A}$ will be preserved well by $\widetilde{\mathcal{A}}$ unless many objects having very close (smaller than the gap defined by $(\varepsilon, \alpha)$) aggregate scores on some query interval; and this is unlikely in real datasets when users choose small values of $(\varepsilon, \alpha)$.

Appendix (Section 10) shows that an algorithm $G$ satisfying Definition 1 implies an algorithm $R$ satisfying Definition 2. That said, for either BREAKPOINTS1 or BREAKPOINTS2, QUERY1 is an $(\varepsilon, 1)$-approximation for $\sigma_i(t_1, t_2)$ and $\mathcal{A}(k, t_1, t_2)$; QUERY2 is an $(\varepsilon, 2\log r)$-approximation for $\sigma_i(t_1, t_2)$ and $\mathcal{A}(k, t_1, t_2)$. Despite the reduction in guaranteed accuracy for QUERY2, in practice its accuracy is not much worse than QUERY1, and it is 1-2 orders of magnitude better in space and construction time; and QUERY1 improves upon EXACT3, the best exact method.

## 3.1 Breakpoints

Our key insight is that $\sigma_i(t_1, t_2)$ does not depend on the number of segments between the boundary times $t_1$ and $t_2$; it only depends on the aggregate $\sigma$ applied to that range. So to approximate the aggregate score of any object within a range, we can discretize them based on the accumulated $\sigma$ value. Specifically, we ensure between no two consecutive breakpoints in $b_j, b_{j+1} \in \mathcal{B}$ does the value $\sigma_i(b_j, b_{j+1})$ become too large for an object. Both sets of breakpoints $\mathcal{B}_1$ for BREAKPOINTS1 and $\mathcal{B}_2$ for BREAKPOINTS2 start with $b_0 = 0$ and end with $b_r = T$. Given $b_0$, they sweep forward in time, always constructing $b_j$ before $b_{j+1}$, and define:

$$b_{j+1} \text{ so} \begin{cases} \sum_{i=1}^{m} \sigma_i(b_j, b_{j+1}) = \varepsilon M, & \text{in BREAKPOINTS1}, \\ \max_{i=1}^{m} \sigma_i(b_j, b_{j+1}) = \varepsilon M, & \text{in BREAKPOINTS2}, \end{cases}$$

where $M = \sum_{i=1}^{m} \sigma_i(0, T)$. Note that these breakpoints $b_j$ are not restricted to, and in general will not, occur at the end points of segments of some $o_i$.

Since the total aggregate $\sum_{i=1}^{m} \sigma_i(0, T) = M$, for BREAKPOINTS1 there will be exactly $r = \lceil 1/\varepsilon + 1 \rceil$ breakpoints as each (except for the last $b_r$) accounts for $\varepsilon M$ towards the total integral. For ease of exposition we will assume that $1/\varepsilon$ is integral and drop the $\lceil \cdot \rceil$ notation, hence $1/\varepsilon \cdot \varepsilon M = M$. Next we notice that BREAKPOINTS2 will have at most as many breakpoints as BREAKPOINTS1 since $\max_{i=1}^{m} X_i \leq \sum_{i=1}^{m} X_i$ for any set of $X_i > 0$. However, the inequality is not strict and these quantities could be equal; this implies the two cases could have the same number of breakpoints. This is restricted to the special case where between *every* consecutive pair $b_j, b_{j+1} \in \mathcal{B}$ *exactly one* object $o_i$ has $\sigma_i(b_j, b_{j+1}) = \varepsilon M$ and for *every* other object $o_{i'}$ for $i \neq i'$ has zero aggregate $\sigma_{i'}(b_j, b_{j+1}) = 0$. As we will demonstrate on real data in Section 5 in most reasonable cases the size of BREAKPOINTS2 is dramatically smaller than the size of BREAKPOINTS1.

**Construction of** BREAKPOINTS1**.** We first need to preprocess all of the objects according to individual tuples for each vertex between two line segments. Consider two line segments $s_1$ and $s_2$ that together span from time $t_L$ to time $t_R$ and transition at time $t_M$. If they are part of object $o_i$ then they have values $v_L = g_i(t_L)$, $v_M = g_i(t_M)$, and $v_R = g_i(t_R)$. Then for the vertex at $(t_M, v_M)$ we store the tuple $(t_L, t_M, t_R, v_L, v_M, v_R)$. Then we sort all tuples across all objects according to $t_M$ in ascending order and place them in a queue $Q$. The breakpoints $\mathcal{B}_1$ will be constructed by popping elements from $Q$.

We need to maintain some auxiliary information while processing each tuple. For each tuple, we can compute the slope of its two adjacent segments as $w_L = (v_M - v_L)/(t_M - t_L)$ and $w_R = (v_R - v_M)/(t_R - t_M)$. Between each pair of segment boundaries the value of an object $g_i(t)$ varies linearly according to the slope $w_{i,\ell}$ in segment $g_{i,\ell}$. Thus the sum $\sum_{i=1}^{m} g_i(t)$ varies linearly according to $W(t) = \sum_{i=1}^{m} w_{i,\ell_i}$ if each $i$th object is currently represented by segment $g_{i,\ell_i}$. Also, at any time $t$ we can write the summed value as $V(t) = \sum_{i=1}^{m} g_i(t)$. Now for any two time points $t_1$ and $t_2$ such that no segments starts or ends in the range $(t_1, t_2)$, and given $V(t_1)$ and $W(t_1)$ we can calculate in constant time the sum $\sum_{i=1}^{m} \sigma_i(t_1, t_2) = \frac{1}{2} W(t_1)(t_2 - t_1)^2 + V(t_1)(t_2 - t_1)$. Thus we always maintain $V(t)$ and $W(t)$ for the current $t$.

Since $b_0 = 0$, to construct $\mathcal{B}_1$ we only need to show how to construct $b_{j+1}$ given $b_j$. Starting at $b_j$ we reset to 0 a running sum up to a time $t \geq b_j$ written $I(t) = \sum_{i=1}^{m} \sigma_i(b_j, t)$. Then we pop a tuple $(t_L, t_M, t_R, v_L, v_M, v_R)$ from $Q$ and process it as follows. We update the running sum to time $t_M$ as $I(t_M) = I(t) + \frac{1}{2} W(t)(t_M - t)^2 + V(t)(t_M - t)$. If $I(t_M) < \varepsilon M$, then we update $V(t_M) = V(t) + W(t)(t_M - t)$, then $W(t_M) = W(t) - w_L + w_R$, and pop the next tuple off of $Q$.

If $I(t_M) \geq \varepsilon M$, that means that the break point $b_{j+1}$ occurred somewhere between $t$ and $t_M$. We can solve for this time $b_{j+1}$ in the equation $I(b_{j+1}) = \varepsilon M$ as

$$b_{j+1} = t + \frac{V(t)}{W(t)} + \frac{1}{W(t)} \sqrt{(V(t))^2 - 2W(t)(I(t) - \varepsilon M)}.$$

The slope $W(t)$ has not changed, but we have to update $V(b_{j+1}) = V(t) + W(t) \cdot (b_{j+1} - t)$. Now we reinsert the tuple at the top of $Q$ to begin the process of finding $b_{j+2}$. Since each of $N$ tuples is processed in linear time, the construction time is dominated by the $O((N/B) \log_B N)$ IOs for sorting the tuples.

**Baseline construction of** BREAKPOINTS2**.** While construction of BREAKPOINTS1 reduces to a simple scan over all segments (represented as tuples), computing BREAKPOINTS2 is not as easy because of the replacement of the sum operation with a max. The difficulties come in resetting the maintained data at each breakpoint.

Again, we first need to preprocess all of the objects according to individual tuples for each line segment. We store the $\ell$th segment of $o_i$ as the tuple $s_{i,\ell} = (t_L, t_R, v_L, v_R, i)$ which stores the left and right endpoints of the segment in time as $t_L$ and $t_R$, respectively, and also stores the values it has at those times as $v_L = g_i(t_L)$ and $v_R = g_i(t_R)$, respectively. Note for each segment $s_{i,\ell}$ we can compute its slope $w_{i,\ell} = (v_R - v_L)/(t_R - t_L)$. Then we sort all tuples across all objects according to $t_L$ in ascending order and place them in a queue $Q$. The breakpoints $\mathcal{B}_2$ will be constructed by popping elements from $Q$.

By starting with $b_0 = 0$, we only need to show how to compute $b_{j+1}$ given $b_j$. We maintain a running integral $I_i(t) = \sigma_i(b_j, t)$ for each object. Thus at the start of a new break point $b_j$, each integral is set to 0. Then for each new segment $s_{i,\ell}$ that we pop from $Q$, we update $I_i(t)$ to $I_i(t_R) = I_i(t) + (v_R - v_L)(t_R - t_L)/2$. If $I_i(t_R) < \varepsilon M$, then we pop the next tuple from $Q$ and continue.

However, if the updated $I_i(t_R) \geq \varepsilon M$, then it means we have an event before the next segment will be processed from $o_i$. As before with BREAKPOINTS1, we calculate $\hat{b}_{j+1,i} = t + \frac{g_i(t)}{w_{i,\ell}} + \frac{1}{w_{i,\ell}} \sqrt{(g_i(t))^2 - 2w_{i,\ell}(I_i(t) - \varepsilon M)}$. This is not necessarily the location of the next breakpoint $b_{j+1}$, but if the breakpoint is caused by $o_i$, then this will be it. We call such objects for which we have calculated $\hat{b}_{j+1,i}$ as *dangerous*. We let $\hat{b}_{j+1} = \min \hat{b}_{j+1,i}$ (where $\hat{b}_{j+1,i}$ is implicitly $\infty$ if it is not dangerous). To determine the true next breakpoint we keep popping tuples from $Q$ until for the current tuple $t_L > \hat{b}_{j+1}$. This indicates no more segment endpoints occur before some object $o_i$ reaches $I_i(t) = \varepsilon M$. So we set $b_{j+1} = \hat{b}_{j+1}$, and reset maintained values in preparation for finding $b_{j+2}$.

Assuming $\Omega(m/B)$ internal memory space, this method runs in $O((N/B) \log_B N)$ IOs, as we can maintain $m$ running sums in memory. We can remove this assumption in $O((N/B) \log_B N)$ IOs with some technical tricks which we omit the details of for space. To summarize, after sorting in $O(\log_B N)$ passes on the data, we determine for each segment from each $o_i$ how many segments occur again before another segment from $o_i$ is seen. We then keep the auxiliary information for each object (e.g. running sums) in an IO-efficient priority queue [5] on the objects sorted by the order in which a segment from each object will next appear.

However, with limited internal space or in counting internal runtime, this method is still potentially slower than finding BREAKPOINTS1 since it needs to reset each $I_i(b_{j+1}) = 0$ when we reach a new breakpoint. This becomes clear when studied from an internal memory runtime perspective, where this method may take $O(rm + N \log N)$ time.

**Efficient construction of** BREAKPOINTS2**.** We can avoid the extra $O(rm)$ term in the run time by using clever bookkeeping that ensures we do not have to reset too much each time we find a break point. Appendix in Section 9.1 of our technical report [10] shows:

**Lemma 1** BREAKPOINTS2 *can be built in* $O(N \log N)$ *time (for* $N > 1/\varepsilon$*). Its size is* $r = O(1/\varepsilon)$*; and it takes* $O((N/B) \log_B N)$ *IOs to construct.*

**Remarks.** For specific datasets there may be other specialized ways of choosing breakpoints. For real world datasets, such as the MesoWest data as shown in Figure 1, our methods are both efficient and have excellent approximation quality (see Section 5).

## 3.2 Index Breakpoints and Queries

Given a set of breakpoints $\mathcal{B}$ (either $\mathcal{B}_1$ or $\mathcal{B}_2$), we show how to answer queries on the full dataset approximately. The approximation guarantees are based on the following property that holds for BREAKPOINTS1 $\mathcal{B}_1$ and BREAKPOINTS2 $\mathcal{B}_2$. For any query interval $(t_1, t_2)$, let $(\mathcal{B}(t_1), \mathcal{B}(t_2))$ be the associated *approximate interval*, where $\mathcal{B}(t_1)$ (resp. $\mathcal{B}(t_2)$) is the smallest breakpoints in $\mathcal{B}$ such that $\mathcal{B}(t_1) \geq t_1$ (resp. $\mathcal{B}(t_2) \geq t_2$); see Figure 8.

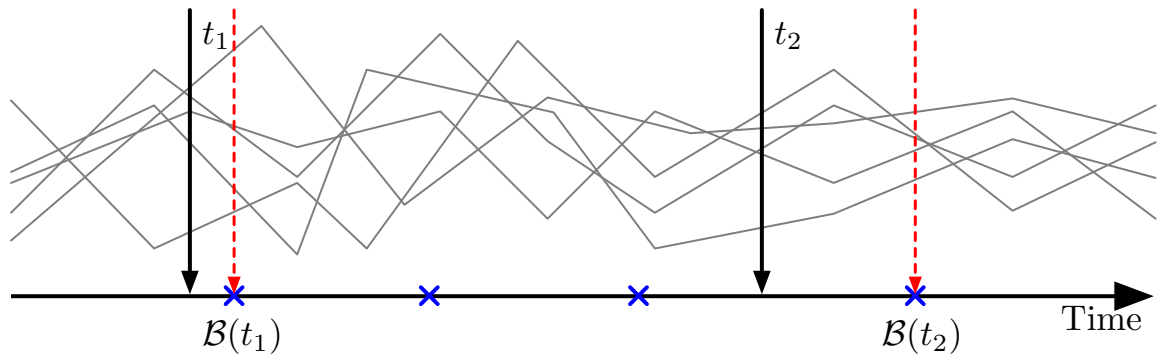

**Figure 8: Associated approximate interval.**

**Lemma 2** *For any query* $[t_1, t_2]$ *and associated approximate interval* $[\mathcal{B}(t_1), \mathcal{B}(t_2)]$*:* $\forall o_i$, $|\sigma_i(t_1, t_2) - \sigma_i(\mathcal{B}(t_1), \mathcal{B}(t_2))| \leq \varepsilon M$.

PROOF. Both $\mathcal{B}_1$ and $\mathcal{B}_2$ guarantee that between any two consecutive breakpoints $b_j, b_{j+1} \in \mathcal{B}$ that for any object $\sigma_i(b_j, b_{j+1}) \leq \varepsilon M$. This property is guaranteed directly for BREAKPOINTS2, and is implied by BREAKPOINTS1 because for any object $o_i$ it holds that $\sigma_i(t_1, t_2) \leq \sum_{j=1}^{m} \sigma_j(t_1, t_2)$ for each $\sigma_j(t_1, t_2) \geq 0$, which is the case since we assume positive scores (this restriction is removed in Section 4).

Hence, by changing the query interval from $[t_1, t_2]$ to $[\mathcal{B}(t_1), t_2]$ the aggregate can only *decrease*, and can decrease by at most $\varepsilon M$. Also, by changing the interval from $[\mathcal{B}(t_1), t_2]$ to $[\mathcal{B}(t_1), \mathcal{B}(t_2)]$ the aggregate can only *increase*, and can increase by at most $\varepsilon M$. Thus the inequality holds since each endpoint change can either increase *or* decrease the aggregate by at most $\varepsilon M$. □

We now present two query methods, and associate data structures, called QUERY1 and QUERY2.

**Nested B+-tree queries.** For QUERY1 we consider all $\binom{r}{2}$ intervals with a breakpoint from $\mathcal{B}$ at each endpoint. For each of these intervals $[b_j, b_{j'}]$, we construct the $k_{\max}$ objects with the largest aggregate $\sigma_i(b_j, b_{j'})$. Now we can show that this nested B+-tree yields an $(\varepsilon, 1)$-approximation for both the aggregate scores and $\mathcal{A}(k, t_1, t_2)$ for any $k \leq k_{\max}$.

To construct the set of $k_{\max}$ objects associated with each interval $[b_j, b_{j'}]$ we use a single linear sweep over all segments using operations similar to EXACT1. Starting at each breakpoint $b_j$, we initiate a running integral for each object to represent the intervals with $b_j$ as their left endpoint. Then at each other breakpoints $b_{j'}$ we output the $k_{\max}$ objects with largest running integrals starting at each $b_j$ up to $b_{j'}$ to represent $[b_j, b_{j'}]$. That is, we maintain $O(r)$ sets of $m$ running integrals, one for each left break point $b_j$ we have seen so far (to avoid too much internal space in processing all $N$ segments, we use a single IO-efficient priority queue as in constructing BREAKPOINTS2, where each of $m$ objects in the queue now also stores $O(r)$ running sums.) We also maintain $O(r)$ priority queues of size $k_{\max}$ for each left endpoint $b_j$, over each set of $m$ running integrals on different objects. This takes $O((N/B)(\log_B(mr) + r \log_B k_{\max}) + r(rk_{\max}/B + 1))$ IOs, where the last item counts for the output size (since we have $O(r^2)$ intervals and each interval stores $k_{\max}$ objects). We assume $rk_{\max} < N$ (to simplify and so index size $O(r^2 k_{\max})$ is feasible); hence, the last term is absorbed in $O(\cdot)$.

To index the set of these intervals, we use nested set of B+-trees. We first build a B+-tree $T_{top}$ on the breakpoints $\mathcal{B}$. Then for each leaf node associated with $b_j$, we point to another B+-tree $T_j$ on $\mathcal{B}'_j$, where $\mathcal{B}'_j = \{b \in \mathcal{B} \mid b > b_j\}$. The top level B+-tree $T_{top}$ indexes the left endpoint of an interval $[b_j, b_{j'}]$ and the lower level B+-tree $T_j$ pointed to by $b_j$ in $T_{top}$ indexes the right end point $b_{j'}$ (for all $b_{j'} > b_j$). We build $O(r)$ B+-trees of size $O(r)$, hence, this step takes $O(r^2/B)$ IOs (by bulkloading). Again, we assume $r^2 < N$, and this cost will also be absorbed in the construction cost.

Now we can query any interval in $O(\log_B r)$ time, since each B+-tree requires $O(\log_B r)$ to query, and for a query top-$k(t_1, t_2, \sigma)$, we use $T_{top}$ to find $\mathcal{B}(t_1)$, and the associated lower level B+-tree of $\mathcal{B}(t_1)$ to find $\mathcal{B}(t_2)$, which gives the top $k_{\max}$ objects in interval $[\mathcal{B}(t_1), \mathcal{B}(t_2)]$. We return the top $k$ objects from them as $\widetilde{\mathcal{A}}$ (see Figure 9). The above and Lemma 2 imply the following results.

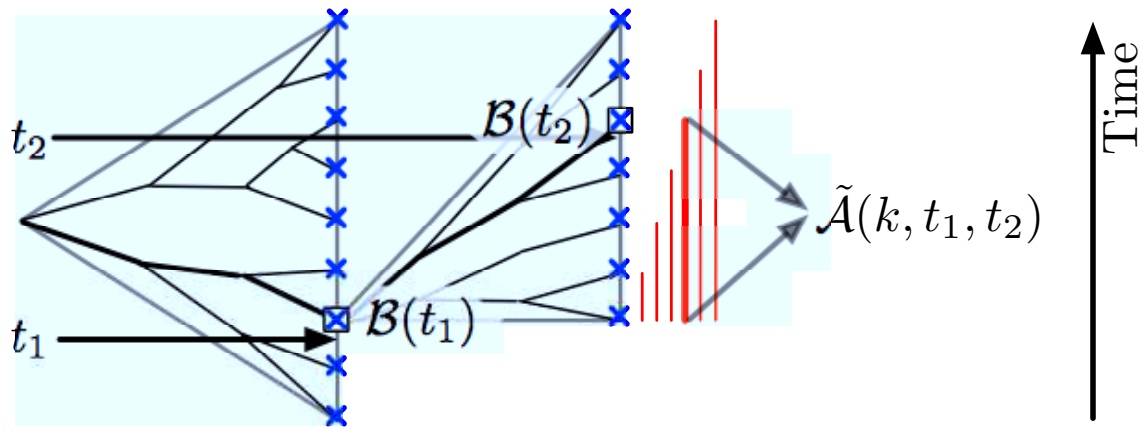

**Figure 9: Illustration of** QUERY1**.**

**Lemma 3** *Given breakpoints* $\mathcal{B}$ *of size* $r$ *(*$r^2 < N$ *and* $rk_{\max} < N$*),* QUERY1 *takes* $O((N/B)\,(\log_B(mr) + r \log_B k_{\max}))$ *IOs to build, has size* $\Theta(r^2 k_{\max}/B)$*, and returns* $(\varepsilon, 1)$*-approximate top-*$k$ *queries, for any* $k \leq k_{\max}$*, in* $O(k/B + \log_B r)$ *IOs.*

**Dyadic interval queries.** QUERY1 provides very efficient queries, but requires $\Omega(r^2 k_{\max}/B)$ blocks of space which for small values of $\varepsilon$ can be too large (as $r = O(1/\varepsilon)$ in both types of breakpoints). For arbitrarily small $\varepsilon$, it could be that $r^2 > N$. It also takes $\Omega(rN \log k_{\max})$ time to build. Thus, we present an alternative approximate query structure, called QUERY2, that uses only $O(rk_{\max}\ /B)$ space, still has efficient query times and high empirical accuracy, but has slightly worse accuracy guarantees. It is a $(\varepsilon, 2\log r)$-approximation for both $\sigma_i(t_1, t_2)$ and $\mathcal{A}(k, t_1, t_2)$.

We consider all dyadic intervals, that is all intervals $[b_j, b_{j'}]$ where $j = h2^\ell + 1$ and $j' = (h+1)2^\ell$ for some integer $0 \leq \ell < \log r$ and $0 \leq h \leq r/2^\ell - 1$. Intuitively, these intervals represent the span of each node in a balanced binary tree. At each level $\ell$ the intervals are of length $2^\ell$, and there are $\lceil r/2^\ell \rceil$ intervals. There are less than $2r + \log r$ such intervals in total since there are $r$ at level 0, $\lceil r/2 \rceil$ at level 1, and so on, geometrically decreasing.

As with QUERY1 for each dyadic interval $[b_j, b_{j'}]$ we find the $k_{max}$ objects with the largest $\sigma_i(b_j, b_{j'})$ in a single sweep over all $N$ segments. There are $\log r$ active dyadic intervals at any time, one at each level, so we maintain $\log r$ running integrals per object. We do so again using two IO-efficient priority queues. One requires $O((1/B) \log_B(m \log r))$ IOs per segment, the elements correspond to objects sorted by which have segments to processes next, and each element stores the $\log r$ associated running integrals.

The second is *a set* of $\log r$ IO-efficient priority queues of size $k_{\max}$, sorted by the value of the running integral; each requires $O((1/B)\log_B k_{\max})$ IOs per segment. The total construction is $O((N/B)(\log_B(m\log r) + \log r \log_B k_{\max}))$ IOs.

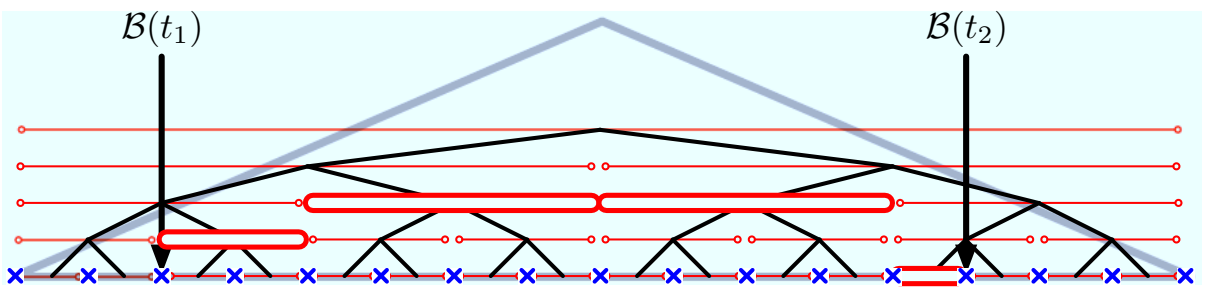

**Figure 10: Illustration of QUERY2.**

In dyadic intervals any interval $[b_1, b_2]$ can be formed as the disjoint union of at most $2\log r$ dyadic intervals. We use this fact as follows: for each query interval $[t_1, t_2]$ we determine the at most $2\log r$ dyadic intervals that decompose the associated approximate query interval $[\mathcal{B}(t_1), \mathcal{B}(t_2)]$. For each such dyadic interval, we retrieve the top-$k$ objects and scores from its associated top-$k_{\max}$ objects ($k \le k_{\max}$), and insert them into a candidate set $\mathcal{K}$, adding scores of objects inserted more than once. The set $\mathcal{K}$ is of size at most $k2\log r$. We return the $k$ objects with the top $k$ summed aggregate scores from $\mathcal{K}$.

**Lemma 4** QUERY2 *$(\varepsilon, 2\log r)$-approximations $\mathcal{A}(k, t_1, t_2)$.*

PROOF. Converting $[t_1, t_2]$ to $[\mathcal{B}(t_1), \mathcal{B}(t_2)]$ creates at most $\varepsilon M$ error between $\sigma_i(t_1, t_2)$ and $\sigma_i(\mathcal{B}(t_1), \mathcal{B}(t_2))$, as argued in Lemma 2. This describes the additive $\varepsilon M$ term in the error, and allows us to hereafter consider only the lower bound on score over the approximate query interval $[\mathcal{B}(t_1), \mathcal{B}(t_2)]$.

The relative $2\log r$ factor is contributed to by the decomposition of $[\mathcal{B}(t_1), \mathcal{B}(t_2)]$ into at most $2\log r$ disjoint intervals. For each object $o_i \in \mathcal{A}(t_1, t_2)$, some such interval $[b_j, b_{j'}]$ must satisfy $\sigma_i(b_j, b_{j'}) \ge \sigma_i(\mathcal{B}(t_1), \mathcal{B}(t_2))/(2\log r)$. For this interval, if $o_i$ is in the top-$k$ then we return a value at least $\sigma_i(b_j, b_{j'}) \ge \sigma_i(\mathcal{B}(t_1), \mathcal{B}(t_2))/(2\log r)$. If $o_i$ is not in the top-$k$ for $[b_j, b_{j'}]$ then each object $o_{i'}$ that is in that top-$k$ set has

$$\sigma_{i'}(\mathcal{B}(t_1), \mathcal{B}(t_2)) \ge \sigma_{i'}(b_j, b_{j'}) \ge \sigma_i(b_j, b_{j'}) \ge \frac{\sigma_i(\mathcal{B}(t_1), \mathcal{B}(t_2))}{2\log r}.$$

Thus, there must be at least $k$ objects $o_{i'} \in \tilde{\mathcal{A}}(\mathcal{B}(t_1), \mathcal{B}(t_2))$ with $\sigma_{i'}(\mathcal{B}(t_1), \mathcal{B}(t_2)) \ge \sigma_i(\mathcal{B}(t_1), \mathcal{B}(t_2))/(2\log r)$. □

To efficiently construct the set $\mathcal{K}$ of at most $k2\log r$ potential objects to consider being in $\tilde{\mathcal{A}}(k, t_1, t_2)$, we build a balanced binary tree over $\mathcal{B}$. Each node (either an internal node or leaf node) corresponds to a dyadic interval (see Figure 10). We construct the set of such intervals that form the disjoint union over $[\mathcal{B}(t_1), \mathcal{B}(t_2)]$ as follows. In phase 1, starting at the root, if $[t_1, t_2]$ is completely contained within one child, we recurse to that child. Phase 2 begins when $[t_1, t_2]$ is split across both children of a node, so we recur on each child. On the next step Phase 3 begins, we describe the process for the left child; the process is symmetric for the right child. If $t_1$ is within the right child, we recur to that child. If $t_1$ is within the left child, we return the dyadic interval associated with right child and recur on the left child. Finally, if $t_1$ separates the left child from the right child, we return the dyadic interval associated with the right child and terminate. Since the height of the tree is at most $\log r$, and we return at most one dyadic interval at each level for the right and left case of phase 3, then there are at most $2\log r$ dyadic intervals returned. The above idea can be easily generalized to a B+-tree (simply with larger fanout) if $r$ is large.

**Lemma 5** *Given breakpoints $\mathcal{B}$ of size $r$,* QUERY2 *requires size $\Theta(rk_{\max}/B)$, takes $O((N/B)(\log_B(m\log r) + \log r \log_B k_{\max}))$ cost to build, and answers $(\varepsilon, 2\log r)$-approximate top-$k$ queries, for any $k \le k_{\max}$, in $O(k\log r \log_B k)$ IOs.*

PROOF. The error bound follows from Lemma 4, and the construction time is argued above. The query time is dominated by maintaining a size $k$ priority queue over the set $\mathcal{K}$ with $O(k\log r)$ objects inserted, from $k$ objects in $O(\log r)$ dyadic intervals. □

## 3.3 Combined Approximate Methods

Finally we formalize different approximate methods: APPX1-B, APPX2-B, APPX1, APPX2. As shown in Figure 7 the methods vary based on how we combine the construction of breakpoints and the query structure on top of them. APPX1 and APPX2 use BREAKPOINTS2 followed by either QUERY1 or QUERY2, respectively. As we will demonstrate in Section 5, BREAKPOINTS2 is superior to BREAKPOINTS1 in practice; so, we designate APPX1-B (BREAKPOINTS1 +QUERY1) the *basic* version of APPX1, and APPX2-B (BREAKPOINTS1 +QUERY2) the basic version of APPX2.

The analysis between the basic and improved versions are largely similar, hence, we only list the improved versions in Table 3. In particular, for the below results, since $r = \Theta(1/\varepsilon)$ in BREAKPOINTS1, we can replace $r$ with $1/\varepsilon$ for the basic results.

APPX1 computes $r = O(1/\varepsilon)$ breakpoints $\mathcal{B}_2$ using BREAKPOINTS2 in $O((N/B)\log_B(N/B))$ IOs. Then QUERY1 requires $O(r^2 k_{\max}/B)$ space, $O((N/B)(\log_B(mr) + r\log_B k_{\max}))$ construction IOs, and can answer $(\varepsilon, 1)$-approximate queries in $O(k/B + \log_B r)$ IOs. Since $m, r < N$, this simplifies the total construction IOs to $O((N/B)(\log_B N + r\log_B k_{\max})$, the index size to $O(r^2 k_{\max}/B)$ and the IOs for an $(\varepsilon, 1)$-approximate top-$k$ query to $O(k/B + \log_B r)$.

In APPX2, QUERY2 has $O(rk_{\max}/B)$ space, builds in $O((N/B)(\log_B(m\log r) + \log r \log_B k_{\max}))$ IOs, and answers $(\varepsilon, 2\log r)$-approximate queries in $O(k\log r \log_B k)$ IOs. As $m, r < N$, the bounds simplify to $O((N/B)\ (\log_B N + \log r \log_B k_{\max}))$ build cost, $O(k\log r \log_B k)$ query IOs, and $O(rk_{\max}/B)$ index size. We also consider a variant APPX2+, which discovers the exact aggregate value for each object in $\mathcal{K}$ using a B+-tree from EXACT2. This increases the index size by $O(N/B)$ (basically just storing the full data), and increases the query IOs to $O(k\log r \log_B k)$, but significantly improves the empirical query accuracy.

# 4. OTHER REMARKS

**Updates**. In most applications, temporal data receive updates only at the current time instance, which extend a temporal object for some specified time period. In this case, we can model an update to an object $o_i$ as appending a new line segment $g_{i,n_i+1}$ to the end of $g_i$, where that $g_{i,n_i+1}$'s left end-point is $(t_{i,n_i}, v_{i,n_i})$ (the right end-point of $g_{i,n_i}$); $g_{i,n_i+1}$'s right end-point is $(t_{i,n_i+1}, v_{i,n_i+1})$.

Handling updates in exact methods are straightforward. In EXACT1, we insert a new entry $(t_{i,n_i}, g_{i,n_i+1})$ into the B+-tree; hence the update cost is $O(\log_B N)$ IOs. In EXACT2, we insert a new entry $(t_{i,n_i+1}, (g_{i,n_i+1}, \sigma_i(I_{i,n_i+1}))$ to the B+-tree $T_i$, where $I_{i,n_i+1} = [t_{i,0}, t_{i,n_i+1}]$. We can compute $\sigma_i(I_{i,n_i+1})$ based on $\sigma_i(I_{i,n_i})$ and $g_{i,n_i+1}$ in $O(1)$ cost; and $\sigma_i(I_{i,n_i})$ is retrieved from the last entry in $T_i$ in $O(\log_B n_i)$ IOs. So, the update cost is $O(\log_B n_i)$ IOs. In EXACT3, a new entry $([t_{i,n_i}, t_{i,n_i+1}], (g_{i,n_i+1}, \sigma_i(I_{i,n_i+1})))$ is inserted into the interval tree $S$. For similar arguments, $\sigma_i(I_{i,n_i})$ is retrieved from $S$ in $O(\log_B N)$ IOs; and then $\sigma_i(I_{i,n_i+1})$ is computed in $O(1)$. The insertion into $S$ is $O(\log_B N)$ IOs [4]. Thus the total update is $O(\log_B N)$ IOs.

Handling updates in approximate methods is more complicated. As such, we described amortized analysis for updates. This approach can be de-amortized using standard technical tricks. The construction of breakpoints depends on a threshold $\tau = \varepsilon M$; however, $M$ increases with updates. We handle this by always constructing breakpoints (and the index structures on top of them) us-ing a fixed value of $\tau$, and when $M$ doubles, we rebuild the structures. For this to work, we assume that it takes $\Omega(N)$ segments before $M$ doubles; otherwise, a segment $\ell$ could have an aggregate of $M/2$, and one has to rebuild the entire query structure immediately after seeing $\ell$. Thus in an amortized sense, we can amortize the construction time $C(N)$ over $\Omega(N)$ segments, and charge $O(C(N)/N)$ to the update time of a segment.

We also need to maintain a query structure and set of breakpoints on top of the segments just added. Adding the breakpoints can be done by maintaining the same IO-efficient data structures as in their initial construction, using $O(\frac{1}{B}\log_B N)$ IOs per segment. To maintain the query structures, we again maintain the same auxiliary variables and running integrals as in the construction. Again, assuming that there are $\Omega(N/r)$ segments between any pair of breakpoints, we can amortize the building of the query structures to the construction cost divided by $N$. The amortized reconstruction or incremental construction of the query structures dominate the cost. For APPX1 we need $O(\frac{1}{B}(\log_B N + r\log_B k_{\max}))$ IOs to update QUERY1. For APPX2 we need $O(\frac{1}{B}(\log_B N + \log r \log_B k_{\max}))$ IOs to update QUERY2.

**General time series with arbitrary functions.** In some time series data, objects are described by arbitrary functions $f$, instead of piecewise linear functions $g$. However, as we explained in Section 1, a lot of efforts have been devoted to approximate an arbitrary function $f$ using a piecewise linear function $g$ in general time series (see [17] and references therein). Furthermore, to understand the flexibility of our methods, it is important to observe that all of our methods also naturally work with any *piecewise polynomial functions* $p$: the only change is that we need to deal with polynomial curve segments, instead of linear line segments. This only affects, in all our methods, how to compute $\sigma_i(I)$ of an interval $I$, which is a subinterval of the interval defined by the two end-points of a polynomial curve segment $p_{i,j}$ (the $j$th polynomial function in the $i$th object). But this can be easily fixed. Instead of using (1) based on a trapezoid, we simply compute it using the integral over $p_{i,j}$, i.e., $\sigma_i(I) = \int_{t\in I} p_{i,j}(t)d(t)$. Given that $p_{i,j}(t)$ is a polynomial function, this can be easily computed. That said, when one needs more precision in representing an arbitrary time series, either one can use more line segments in a piecewise linear representation, or one can use a piecewise polynomial representation. All of our methods work in both cases.

**Negative values.** We have assumed positive score values so far. But this restriction can be easily removed. Clearly, it does not affect our exact methods at all. In the approximate methods, when computing the breakpoints (in either approach), we use the absolute values instead to define $M$ and when searching for a breakpoint. We omit technical details due to the space constraint, but we can show that doing so will still guarantee the same approximations.

**Other aggregates.** Our work focuses on the sum aggregation. This automatically implies the support to the avg aggregation, and many other aggregations that can be expressed as linear combinations of the sum (such as $F_2$, the 2nd frequency moment). However, ranking by some holistic aggregates is hard. An important one in this class is the quantile (median is a special case of the quantile). We leave the question of how to rank large temporal data using some of the holistic aggregates (e.g., quantile) as an open problem.

# 5. EXPERIMENTS

We design all of our algorithms to efficiently consider disk IOs; in particular, we implemented all our methods using the TPIE-library in C++ [2]. This allows our methods to scale gracefully to massive data that does not fit in memory. All experiments were performed on a Linux machine with an Intel Core i7-2600 3.4GHz CPU, 8GB of memory, and a 1TB hard drive.

**Datasets.** We used two large real datasets. The first dataset is a temperature dataset, *Temp*, from the MesoWest project [8]. It contains temperature measurements from Jan 1997 to Oct 2011 from 26,383 distinct stations across the United States. There are almost $N$=2.6 billion total readings from all stations with an average of 98,425 readings per station. For our experiments, we preprocessed the *Temp* dataset to treat *each year of readings from a distinct station as a distinct object*. By aligning readings in this manner we can ask which $k$ stations had the highest aggregate temperatures in a (same) time interval amongst any of the recorded years. After preprocessing, *Temp* has $m$=145,628 objects with an average number of readings per object of $n_{\text{avg}}$=17,833. In each object, we connect all consecutive readings to obtain a piecewise-linear representation.

The second real dataset, *Meme*, was obtained from the Memetracker project. It tracks popular quotes and phrases which appear from various sources on the internet. The goal is to analyze how different quotes and phrases compete for coverage every day and how some quickly fade out of use while others persist for long periods of time. A record has 4 attributes, the URL of the website containing the memes, the time Memetracker observed the memes, a list of the observed memes, and links accessible from the website. We preprocess the *Meme* dataset, converting each record to have a distinct 4-byte integer *id* to represent the URL, an 8-byte double to represent the *time* of the record, and an 8-byte double to represent a record's *score*. A record's score is the number of memes appearing on the website, i.e. it is the cardinality of the list of memes. After preprocessing, *Meme* has almost $m$=1.5 million distinct objects (the distinct URLs) with $N$=100 million total records, an average of $n_{\text{avg}}$=67 records per object. For each object, we connect every two of its consecutive records in time (according to the date) to create a piecewise linear representation of its score.

**Setup.** We use *Temp* as the default dataset. To test the impact of different variables, we have sampled subsets of *Temp* to create datasets of different number of objects ($m$), different number of average line segments per object ($n_{\text{avg}}$, by limiting the maximum value $T$). By default, $m = 50,000$ and $n_{\text{avg}} = 1,000$ in *Temp*, so all exact methods can finish in reasonable amount of time. Still, there are a total of $N = 50\times10^6$ line segments! The default values of other important variables in our experiments are: $k_{\max} = 200$, $k = 50$, $r = 500$ (number of breakpoints in both BREAKPOINTS1 and BREAKPOINTS2), and $(t_2 - t_1) = 20\%T$. The disk block size in TPIE is set to 4KB. For each query-related result, we generated 100 random queries and report *the average*. Lastly, in all datasets, all line segments are sorted by the time value of their left end-point.

**Number of breakpoints.** We first investigate the effect of the number of breakpoints $r$ on different approximate methods, by changing $r$ from 100 to 1000. Figure 11 shows the preprocessing results and Figure 12 shows the query results. Figure 11(a) indicates that given the same number of breakpoints, the value of the error parameter $\varepsilon$ using BREAKPOINTS2 $\mathcal{B}_2$ is much smaller than that in BREAKPOINTS1 $\mathcal{B}_1$ in practice; this confirms our theoretical analysis, since $r = 1/\varepsilon$ in $\mathcal{B}_1$, but $r = O(1/\varepsilon)$ in $\mathcal{B}_2$. This suggests that $\mathcal{B}_2$ offers *much higher* accuracy than $\mathcal{B}_1$ given the same budget $r$ on real datasets. With 500 breakpoints, $\varepsilon$ in $\mathcal{B}_2$ reduces to almost $10^{-8}$, while it is still 0.02 in $\mathcal{B}_1$. Figure 11(b) shows the build time of $\mathcal{B}_1$ and $\mathcal{B}_2$. Clearly, building $\mathcal{B}_1$ is independent to $r$ since its cost is dominated by the linear sweeping of all line segments. The baseline method for building $\mathcal{B}_2$, BREAKPOINTS2-B clearly has a linear dependency on $r$ (on $m$ as well, which is not reflected by this experiment). However, our efficient method of building $\mathcal{B}_2$, BreakPoints2-E, has largely removed this dependency on $r$ as shown in Figure 11(b). It also removed the dependency on $m$, though not shown. In what follows, BreakPoints2-E was used by default. Both $\mathcal{B}_1$ and $\mathcal{B}_2$ can be built fairly fast, in only 80 and 100 seconds respectfully when $r = 500$ (over $50 \times 10^6$ segments!).

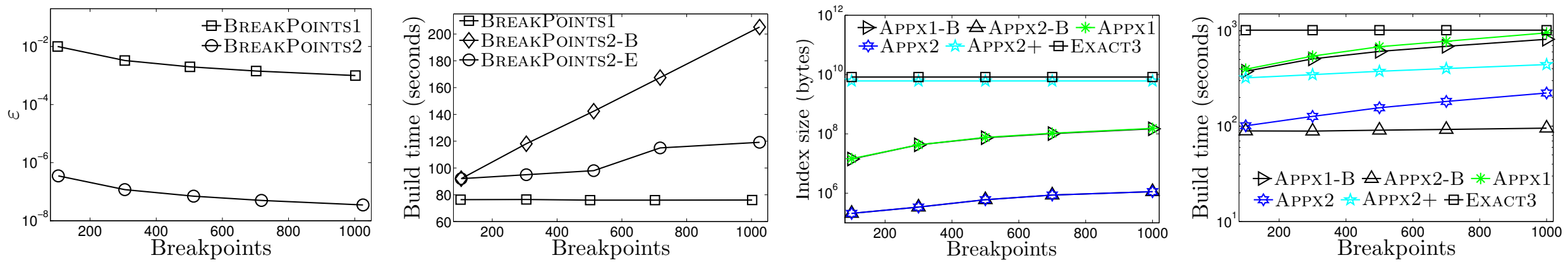

(a) $\varepsilon$ vs $r$. (b) Time vs $r$. (c) Index size. (d) Build time.

**Figure 11: Vary $r$ for approximate methods on Temp: preprocessing results.**

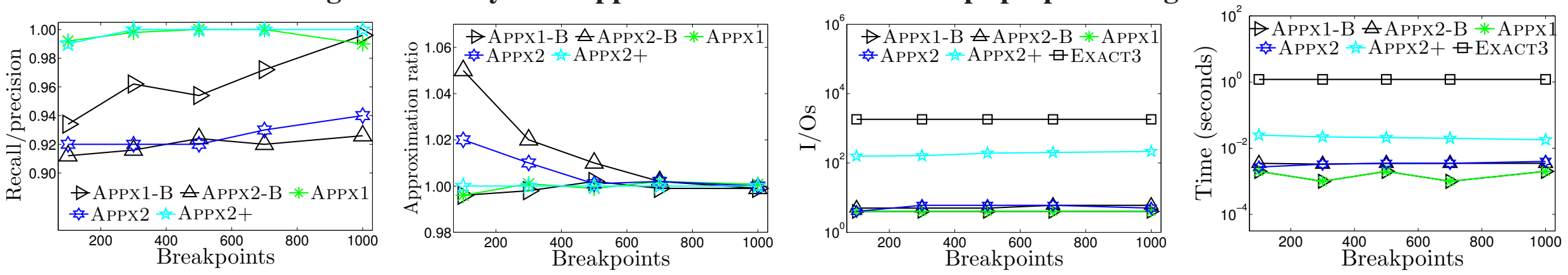

(a) Recall/precision. (b) Ratio. (c) I/Os. (d) Query time.

**Figure 12: Vary $r$ for approximate methods on Temp: query results.**

Next, we investigate the index size and the construction cost of approximate methods, using Exact3 as a reference (as it has the best query performance among all exact methods). Figure 11(c) shows that all approximate methods have much smaller size than Exact3, except Appx2+ which also builds Exact2 since it calculates the exact aggregate score for candidates in $\mathcal{K}$ from Appx2. Clearly, Appx1-B and Appx1 have the same size, basic and improved versions only differ in which types of breakpoints they index using the two-level B+-trees. For the same reason, Appx2-B and Appx2 also have the same size; they index $\mathcal{B}_1$ or $\mathcal{B}_2$ using a binary tree over the dyadic intervals. Appx2-B and Appx2 only have size $O(rk_{\max})$, while Appx1-B and Appx1 have size $O(r^2k_{\max})$ and Exact3 and Appx2+ have linear size $O(N)$, which explains that the size of Appx2-B and Appx2 is more than 2 orders magnitude smaller than the size of Appx1-B and Appx1, which are in turn 3-2 orders magnitude smaller than Exact3 and Appx2+ when $r$ changes from 100 to 1000. In fact, Appx2-B and Appx2 take only 1MB, and Appx1-B and Appx1 take only 100MB, when $r = 1000$; while Exact3 and Appx2+ take more than 10GB. Construction time (for building both breakpoints and subsequent query structures) for approximate methods (including Appx2+) are much faster than Exact3, as shown in Figure 11(d). All structures build in only 100 to 1000 seconds. Not surprisingly, Appx2-B and Appx2 are the fastest, since they only need to find the top $k_{\max}$ objects for $O(r)$ intervals; while Appx1-B and Appx1 need to find the top $k_{\max}$ objects for $O(r^2)$ intervals. Even Appx2+ is significantly faster to build than Exact3 since Exact2 builds faster than Exact3. All approximate methods are generally faster to build than Exact3, by 1-2 orders of magnitude (except for Appx1 when $r$ reaches 1000) since the top $k_{\max}$ objects can be found in a linear sweep over all line segments as explained in Section 3.2.

In terms of the query performance, we first examine the approximation quality of all approximate methods, using both the standard precision/recall (between $\widetilde{\mathcal{A}}$ and $\mathcal{A}$), and the average of the approximation ratios defined as $\widetilde{\sigma}_i(t_1, t_2)/\sigma_i(t_1, t_2)$ for any $o_i$ returned in $\widetilde{\mathcal{A}}$. Since $|\widetilde{\mathcal{A}}|$ and $|\mathcal{A}|$ are both $k$, the precision and the recall will have the same denominator value. Figure 12(a) shows that all approximate methods have precision/recall higher than 90% even in the worst case when $r = 100$; in fact, Appx1 and Appx2+ have precision/recall close to 1 in all cases. Figure 12(b) further shows that Appx1, Appx1-B, and Appx2+ have approximate ratios on the aggregate scores very close to 1, where as Appx2 and Appx2-B have approximation ratios within 5% of 1. In both figures, Appx1 and Appx2 using $\mathcal{B}_2$ are indeed better than their basic versions Appx1-B and Appx2-B using $\mathcal{B}_1$, since given the same number of breakpoints, $\mathcal{B}_2$ results in much smaller $\varepsilon$ values (see Figure 11(a)). Similar results hold for Appx2+, and are omitted to avoid clutter. Nevertheless, all methods perform much better in practice than their theoretical error parameter $\varepsilon$ suggests (which indicates worst-case analysis). Not surprisingly, both types of approximation qualities from all approximate methods improve when $r$ increases; but $r = 500$ already provides excellent qualities.

Finally, in terms of query cost, approximate methods are clear winners over the best exact method Exact3, with better IOs in Figure 12(c) and query time in Figure 12(d). In particular, Appx1-B and Appx1 (reps. Appx2-B and Appx2) have the same IOs given the same $r$ values, since they have identical index structures except different values of entries to index. These four methods have the smallest number of IOs among all methods, in particular, 6-8 IOs in all cases. All require only two queries in a B+-tree of size $r$; a top-level and lower-level tree for Appx1 and Appx1-B, and a left- and right-endpoint query for Appx2 and Appx2-B. Appx2+ is slower with about 100 to 150 IOs in all cases, due to the fact that after identifying the candidate set $\mathcal{K}$, it needs to verify the exact score of each candidate. But, since it only needs to deal with $2k \log r$ candidates in the worst case, and in practice, $|\mathcal{K}| \ll 2k \log r$, its IOs are still very small. In contrast, the best exact method Exact3 takes more than 1000 IOs.

Smaller IO costs lead to much better query performance; all approximate methods outperform the best exact method Exact3 by at least 2 orders of magnitude in Figure 12(d). In particular, they generally take less than 0.01 seconds to answer a top-50($t_1, t_2$, sum) query, in 20% time span over the entire temporal domain, over $50 \times 10^6$ line segments from $50,000$ objects; while the best exact method Exact3 takes around 1 second for the same query. The fastest approximate method only takes close to 0.001 second!

From these results, clearly, Appx1 and Appx2 using $\mathcal{B}_2$ are better than their corresponding basic versions Appx1-B and Appx2-B using $\mathcal{B}_1$, given the same number of breakpoints; and $r = 500$ already gives excellent approximation quality (the same holds for Appx2+, which we omit to avoid clutter). As such, we only use Appx1, Appx2, and Appx2 + for the remaining experiments with

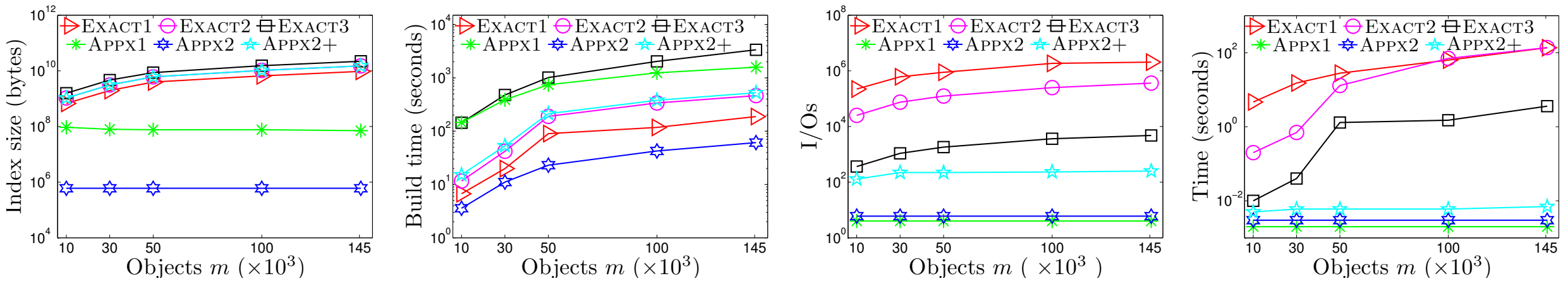

(a) Index size. (b) Build time. (c) Query I/Os. (d) Query time.

**Figure 13: Vary number of objects $m$ on Temp.**

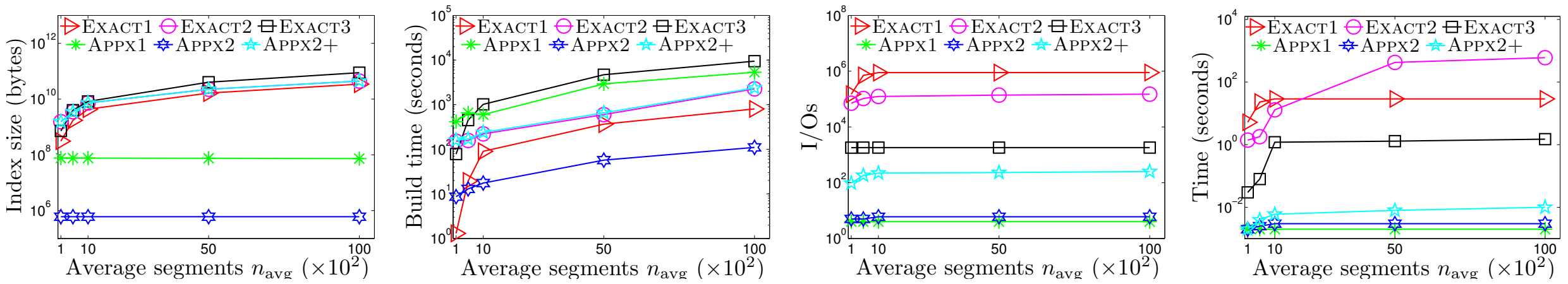

(a) Index size. (b) Build time. (c) Query I/Os. (d) Query time.

**Figure 14: Vary average number of segments $n_{\text{avg}}$ on Temp.**

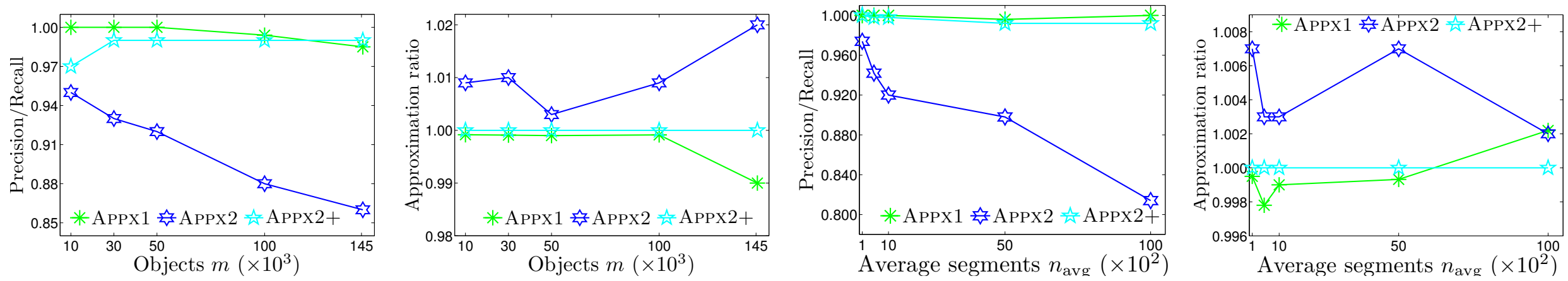

(a) $m$ vs. Precision/Recall. (b) $m$ vs. Ratio. (c) $n_{\text{avg}}$ vs. Precision/Recall. (d) $n_{\text{avg}}$ vs. Ratio.

**Figure 15: $m$ and $n_{\text{avg}}$ vs approximation quality for Temp.**

$r = 500$. Among the three, APPX2+ is larger and slower to build than APPX1, followed by APPX2; the fastest to query are APPX1 and APPX2, then APPX2+; but APPX1 and APPX2+ have better approximation quality than APPX2 (as shown in later experiments and as suggested by their theoretical guarantees for APPX1).

**Scalability.** Next, we investigate the scalability of different methods, using all three exact methods and the three selected approximate methods, when we vary the number of objects $m$, and the average number of line segments per object $n_{\text{avg}}$, in the *Temp* dataset. Figures 13, 14, and 15 show the results. In general, the trends are very consistent and agree with our theoretical analysis. All exact methods consume linear space $O(N)$ and takes $O(N \log N)$ time to build. EXACT3 is clearly the overall best exact method in terms of query costs, outperforming the other two by 2-3 orders of magnitude in terms of IOs and query time (even though it costs slightly more to build). In general, EXACT3 takes hundreds to a few thousand IOs, and about 1 to a few seconds to answer an aggregate top-$k(t_1, t_2, \text{sum})$ query in the *Temp* dataset (with a few hundred million segments from 145,628 objects). Its query performance is not clearly affected by $n_{\text{avg}}$, but has a linear dependency on $m$.

The approximate methods consistently beat the best exact algorithm in query performance by more than 2 orders of magnitude in terms of running time. Even on the largest dataset with few hundred million segments from 145,628 different objects, they still take less than 0.01 seconds per query! Among the three, APPX1 and APPX2 clearly take fewer IOs, since their query cost is actually independent of both $m$ and $n_{\text{avg}}$! APPX2+'s query IO does depend on $\log n_{\text{avg}}$, but is independent of $m$; hence, it is still very small. APPX1 (and even more so APPX2+) occupy much more space, and takes much longer to build. Nevertheless, both APPX1 and APPX2 have much smaller index size than EXACT3, by 4 (APPX1) and 6 (APPX2) orders of magnitude respectively. More importantly, their index size is independent of both $m$ and $n$! In terms of the construction cost, APPX2-B is the most efficient to build (1-2 orders of magnitude faster than all other methods except APPX2).

Figure 15 shows that both APPX1 and APPX2+ retain their high approximation quality when $m$ or $n_{\text{avg}}$ vary; despite some fluctuation, precision/recall and approximation ratios in both APPX1 and APPX2+ stay very close to 1. APPX2 remains at an acceptable level of accuracy, especially considering the index size is 1MB from 50GB of data! Although the precision/recall drops as $n_{\text{avg}}$ and $m$ increases, the very accurate approximation ratio indicates this is because there are many *very* similar objects.

**Query time interval.** Based on our cost analysis, clearly, the length of the query time interval does not affect the query performance of most of our methods, except for EXACT1 that has a linear dependency on $(t_2 - t_1)$ (since it has to scan more line segments). In Figure 16(a) and 16(b) we notice EXACT1 has a linear increase in both I/Os and running time (note the log-scale of the plots) and even for small ($2\%T$) query intervals, it is still much slower than EXACT3 and approximate methods.

In Figures 16(c) and 16(d) we analyze the quality of all approximation techniques as the query interval increases. APPX1 and APPX2+ clearly have the best precision/recall and approximation ratio with a precision/recall above 99% and ratio very close to 1 in all cases. APPX2 shows a slight decline in precision/recall from roughly 98% to above 90% as the size of $(t_2 - t_1)$ increases from 2% to 50% of the maximum temporal value $T$. This decrease in precision/recall is reasonable since as we increase $(t_2 - t_1)$ the number of dyadic intervals which compose the approximate query interval $[\mathcal{B}(t_1), \mathcal{B}(t_2)]$ typically increases. As the number of dyadic intervals increases there is an increased probability that not every candidate in $\mathcal{K}$ will be in the top-$k_{max}$ over each of the dyadic intervals and so APPX2 will be missing some of a candidate's aggregate scores. This can cause an item to be falsely ejected from the top $k$. The effect of missing aggregate scores is clearly seen in Figure 16(d), which shows APPX2's approximation ratio drops slightly as the time range increases.

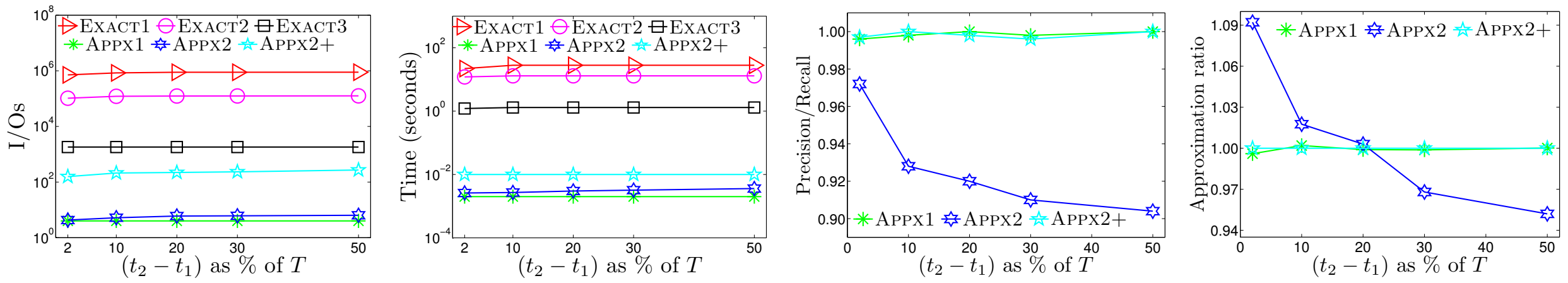

(a) Query I/Os. (b) Query time. (c) Precision/Recall. (d) Ratio.

**Figure 16: Vary size of $(t_2 - t_1)$ as % of $T$ on Temp.**

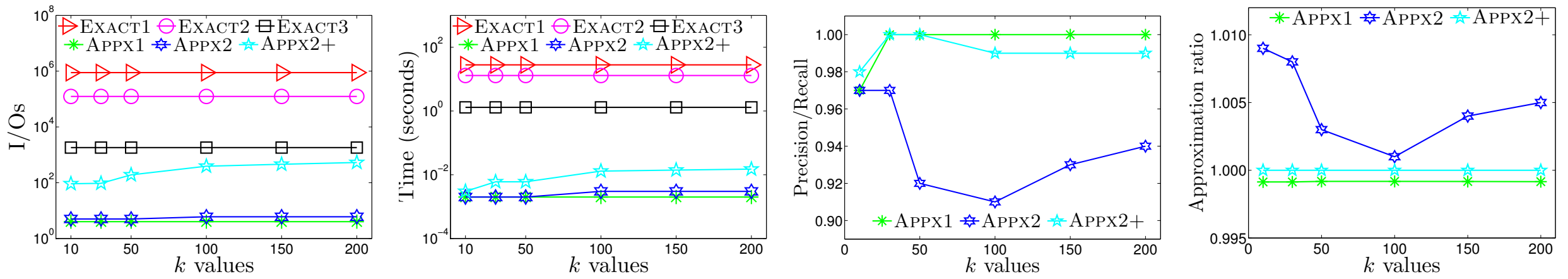

(a) Query I/Os. (b) Query time. (c) Precision/Recall. (d) Ratio.

**Figure 17: Vary $k$ values on Temp.**

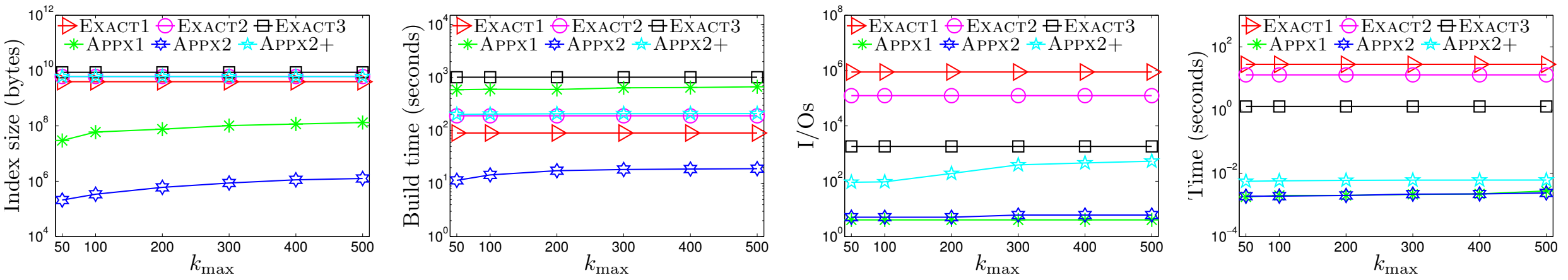

(a) Index size. (b) Construction time. (c) Query IO. (d) Query time.

**Figure 18: Vary $k_{\max}$ on Temp.**

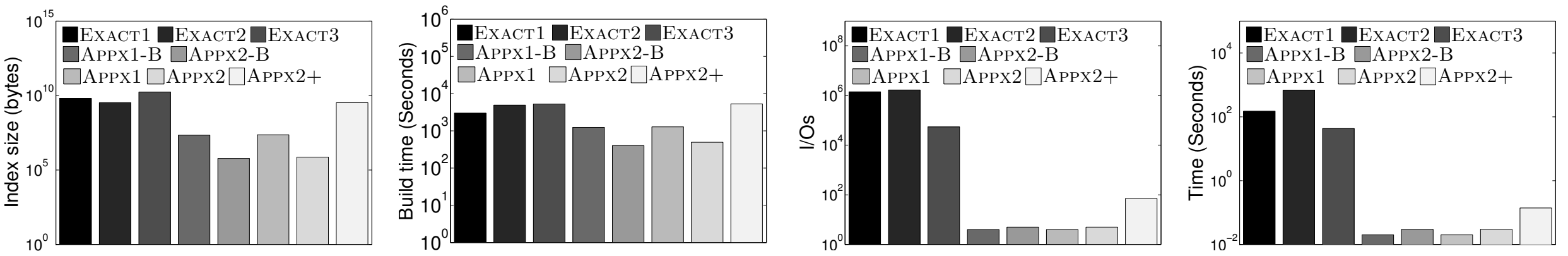

(a) Index size. (b) Build time. (c) I/Os. (d) Query time.

**Figure 19: Meme dataset evaluation.**

**$k$ and $k_{\max}$.** We studied the effect of $k$ and $k_{\max}$; the results are shown in Figures 17 and 18. Figures 17(a) and 17(b) show that the query performance of most methods is not affected by the value of $k$ when it changes from 10 to $k_{\max} = 200$ (a relatively small to moderate change w.r.t. the database size) except for APPX2 and APPX2+. This results since larger $k$ values lead to more candidates in $\mathcal{K}$, which results in higher query cost. Nevertheless, they still have better IOs than the best exact method EXACT3, and much better query cost (still 2 orders of magnitude improvement in the worst case, which can be attributed to the caching effect by the OS). Figure 17(c) and 17(d) show some fluctuation, but no trending changes in accuracy due to variation in $k$.

We vary $k_{\max}$ from 50 to 500 in Figure 18. $k_{\max}$ obviously has no effect on exact methods. It linearly affects the construction cost and the size of index for APPX1 and APPX2, but they are still much better than exact methods even when $k_{\max} = 500$. In terms of query cost, given the same $k$ values, $k_{\max}$ does not clearly affect any approximate methods when it only changes moderately w.r.t. the database size.

**Updates.** As suggested by the cost analysis, the update time for each index structure is roughly proportional to the build time divided by the number of segments. Relative to these build times over $N$, however, EXACT1 is slower because it cannot bulk load, and EXACT2 and APPX2+ are faster because they only update a single B+-tree. For space, we omit these results.

**Meme dataset.** We have also tested all our methods on the *full Meme* dataset (using still $r = 500$ breakpoints for all approximate methods), and the results are shown in Figure 19. In terms of the index size, three exact methods (and APPX2+) are comparable, as seen in Figure 19(a), while other approximate methods take much less space, by 3-5 orders of magnitude! In terms of the construction cost, it is interesting to note that EXACT1 is the fastest to build in this case, due to the bulk-loading algorithm in the B+-tree (since all segments are sorted); while all other methods have some dependency on $m$. But approximate methods (excluding APPX2+) generally are much faster to build than other exact methods as seen in Figure 19(b). They also outperform all exact methods by 3-5 orders of magnitude in IOs in Figure 19(c) and 3-4 orders of magnitude in running time in Figure 19(d). The best exact method for queries is still EXACT3, which is faster than the other two exact methods by 1-2 orders of magnitude. Finally, all approximate methods maintain their high (or acceptable for APPX2) approximation quality on this very bursty dataset, as seen in Figure 20. Note APPX2 achieves this 90% precision/recall and close to 1 approximation ratio while compressing to about 1MB. Also, APPX1 and APPX2 using $\mathcal{B}_2$ show better results than their basic versions APPX1-B and APPX2-B using $\mathcal{B}_1$, given the same number of breakpoints, which agrees with the trend from the *Temp* dataset.

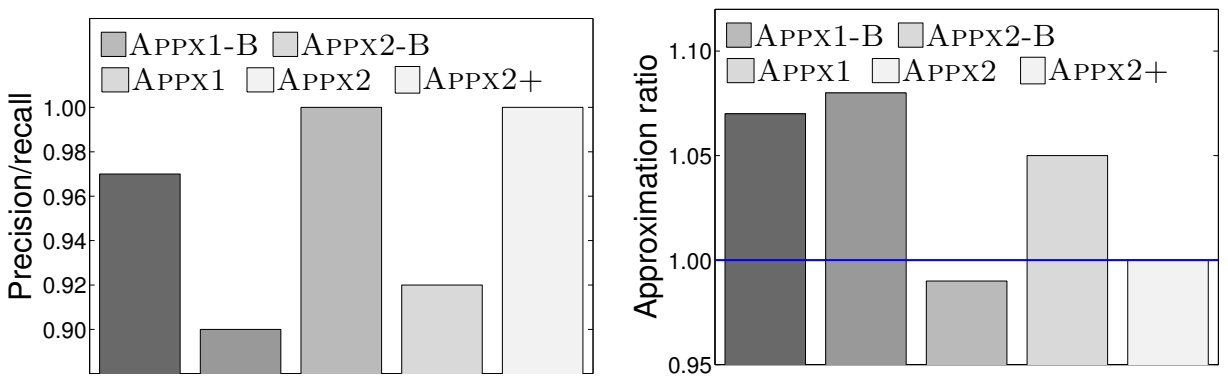

(a) Precision/Recall. (b) Approximation ratio.

**Figure 20: Quality of approximations on Meme.**

# 6. RELATED WORK

To the best of our knowledge, ranking temporal data based on their aggregation scores in a query interval has not been studied before. Ranking temporal data based on the instant top-$k$ definition has been recently studied in [15], however, as we have pointed out in Section 1, one cannot apply their results in our setting. In another work on ranking temporal data [14], they retrieve $k$ objects that are always amongst the top-$k$ list at every time instance over a query time interval. Clearly, this definition is very restrictive and may not even have $k$ objects satisfying this condition in a query interval. This could be relaxed to require an object to be in the top-$k$ list at *most time instances* of an interval, instead of at *all time instances*, like the intuition used in finding durable top-$k$ documents [20], but this has yet to be studied in time series/temporal data. Even then, ranking by aggregation scores still offers quite different semantics, is new, and, is useful in numerous applications.

Our study is related to work on temporal aggregation [22,21]. As mentioned in Section 2, [22,21] focus on multi-versioned keys (instead of time series data), and their objective is to *compute a single aggregation of all keys alive in a query time interval and/or a query key range*, which is different from our definition of aggregation, which is to compute an aggregation over a query time interval, *one per object* (then rank objects based on their aggregation values).

Approximate versions of [22,21] were presented in Tao *et.al.* [18, 19], which also leveraged on a discretization approach (the general principle behind the construction of our breakpoints). As their goal is to approximate aggregates over all keys alive in any query rectangles over the time and the key dimensions (a single aggregate per query rectangle), instead of time-aggregates over each element individually, their approach is not appropriate for our setting.

Our methods require the segmentation of time series data, which has been extensively studied, and the general principles appear in Section 1. A more detailed discussion of this topic is beyond the scope of this work and we refer interested readers to [17,12,16,6,1].

# 7. CONCLUSION

We have presented a comprehensive study on ranking large temporal data using aggregate scores of temporal objects over a query interval which has numerous applications. Our best exact method EXACT3 is much more efficient than baseline methods, and our approximate methods offer further improvements. Interesting open problems include ranking with holistic aggregations (e.g. median and quantiles), and extending to the distributed setting.

# 8. ACKNOWLEDGEMENT

Jeffrey Jestes and Feifei Li were supported in part by NSF Grants IIS-0916488 and IIS-1053979. Feifei Li was also supported in part by a 2011 HP Labs Innovation Research Award. The authors would like to thank Yan Zheng and John Horel from the MesoWest project for providing their datasets and valuable feedback to our study.

# 10. APPENDIX

**Lemma 6** *An algorithm $G$ that satisfies Definition 1 implies an algorithm $R$ that satisfies Definition 2.*

PROOF. $G$ creates $\widetilde{\mathcal{A}}(k, t_1, t_2)$ by finding the top $k$ objects and approximate scores ranked by $\widetilde{\sigma}_i(t_1, t_2)$. By the definition of $G$, $\widetilde{\sigma}_{\widetilde{\mathcal{A}}(j)}(t_1, t_2)$ is an $(\varepsilon, \alpha)$-approximation of $\sigma_{\widetilde{\mathcal{A}}(j)}(t_1, t_2)$. To see $\widetilde{\sigma}_{\widetilde{\mathcal{A}}(j)}(t_1, t_2)$ is an $(\varepsilon, \alpha)$-approximation of $\sigma_{\mathcal{A}(j)}(t_1, t_2)$, note that all $j$ objects $\mathcal{A}(j')$ for $j' \in [0, j]$ satisfy that $\widetilde{\sigma}_{\mathcal{A}(j')}(t_1, t_2) \geq \sigma_{\mathcal{A}(j')}(t_1, t_2)/\alpha - \varepsilon M \geq \sigma_{\mathcal{A}(j)}(t_1, t_2)/\alpha - \varepsilon M$, so $\widetilde{\sigma}_{\widetilde{\mathcal{A}}(j)}(t_1, t_2)$ is at least as large this lower bound. There must be $m - j - 1$ objects $i$ with $\widetilde{\sigma}_i(t_1, t_2) \leq \sigma_{\mathcal{A}(j)}(t_1, t_2) + \varepsilon M$, implying $\widetilde{\sigma}_{\widetilde{\mathcal{A}}(j)}(t_1, t_2) \leq \sigma_{\mathcal{A}(j)}(t_1, t_2) + \varepsilon M$. □